\documentclass{aa}

\usepackage{graphicx}
\usepackage{booktabs}
\usepackage{siunitx}
\usepackage{threeparttable}
\usepackage{txfonts}
\usepackage{lipsum}
\usepackage{subcaption}         
\usepackage{lscape}             
\usepackage{placeins}           

\usepackage{color}
\usepackage{natbib,twoopt}
\usepackage[hyphenbreaks]{breakurl}
\usepackage[breaklinks]{hyperref}      

\bibpunct{(}{)}{;}{a}{}{,}             
\definecolor{cobalt}{rgb}{0.06, 0.2, 0.65}
\hypersetup{
	colorlinks,
	citecolor=cobalt,
	linkcolor=[rgb]{0.8, 0.2, 1.0},
	urlcolor=cobalt,
}
\makeatletter
\newcommandtwoopt{\citeads}[3][][]{\href{http://adsabs.harvard.edu/abs/#3}
	{\def\hyper@linkstart##1##2{}
		\let\hyper@linkend\@empty\citealp[#1][#2]{#3}}}
\newcommandtwoopt{\citepads}[3][][]{\href{http://adsabs.harvard.edu/abs/#3}
	{\def\hyper@linkstart##1##2{}
		\let\hyper@linkend\@empty\citep[#1][#2]{#3}}}
\newcommandtwoopt{\citetads}[3][][]{\href{http://adsabs.harvard.edu/abs/#3}
	{\def\hyper@linkstart##1##2{}
		\let\hyper@linkend\@empty\citet[#1][#2]{#3}}}
\newcommandtwoopt{\citeyearads}[3][][]
{\href{http://adsabs.harvard.edu/abs/#3}
	{\def\hyper@linkstart##1##2{}
		\let\hyper@linkend\@empty\citeyear[#1][#2]{#3}}}
\makeatother

\begin{document}
	
	\title{The role of gas on the collisional runaway path to supermassive black hole seeds}
	
	%
	
	\author{B. Cuevas\inst{1}
		\and B. Reinoso\inst{2}
		\and A. Escala\inst{1}
		\and M. C. Vergara\inst{1,3}
		\and S. Valdebenito\inst{1}
		\and D. R. G. Schleicher\inst{4}
	}
	
	\institute{Departamento de Astronomía, Universidad de Chile, Casilla 36-D, Santiago, Chile
		\and 
		Department of Physics, Gustaf H\"allstr\"omin katu 2, FI-00014, University of Helsinki, Finland \\
		\email{bastian.reinoso@helsinki.fi}
		\and
		Departamento de Astronomía, Facultad de Ciencias Físicas y Matemáticas, Universidad de Concepción, Concepción, Chile
		\and
		Dipartimento di Fisica, Sapienza, Università di Roma, Piazza le Aldo Moro 5, 00185 Roma, Italy
	}
	
	\date{Received August 27, 2026}
	
	
	\abstract
	{The origin of supermassive black holes (SMBHs) remains unclear, despite observational evidence for their rapid formation in the early Universe, as evidenced by active galactic nuclei (AGN) at high redshift.}
	{In this paper, we investigate stellar dynamics in dense, gas-embedded clusters. We explore how gas influences the runaway collision process and the subsequent formation of SMBH seeds in a global instability scenario.} 
	{We perform a series of coupled N-body and SPH simulations using the AMUSE framework, with the stellar and gaseous components interacting gravitationally via the \textsc{BRIDGE} method. We use 12 initial condition models with equal-mass stars following Plummer distributions, with total masses ranging from $10^3$ to $5 \times 10^5~M_{\odot}$, virial radii from $0.005$ to $0.1$ pc, and stellar masses of $1$, $10$ and $50$ $M_{\odot}$.  For each model, we perform four simulations with gas fractions ranging from 0\% to 20\%, resulting in a total of 48 simulations. The gas is sampled with the same spatial distribution as the stars, with a mass ratio per particle of $m_{\rm SPH}/m = 0.01$, where $m$ is the stellar mass.}
	{We find that in most of our simulations, higher gas fractions lead to the formation of more massive black hole, as the core collapses earlier, triggering collisions through gas--star interactions. Our results show increases in BH mass of $\sim 8\%$ for low gas fractions, to $\sim 44\%$ at 20\% gas fraction. The core collapse time is also shortened, with the strongest reduction at $20\%$ gas fraction ($\sim 27\%$).}
	{The transonic regime of our initial conditions is where dynamical friction is most efficient in principle. Although our simulations do not fully resolve this effect, gas--star interactions are the principal mechanism driving the enhancement in BH formation efficiency, in the absence of other dissipation mechanisms (e.g., supernovae, stellar feedback, cooling). This implies that the presence of gas during runaway collisions can lead to the formation of more massive SMBH seeds.}
	
	\keywords{Galaxies: star clusters: general --
		Gas dynamics --
		Hydrodynamics --
		Methods: numerical --
		Stars: black holes --
		Stars: kinematics and dynamics
	}
	
	\maketitle
	\nolinenumbers
	\section{Introduction}
	\label{sec:Introduction}
	The origin of supermassive black holes (SMBHs) is currently a topic of active research, and the formation mechanism remains unclear \citep{woods2019titans}. There is evidence of SMBHs with masses in the range $\sim 10^{6}$--$10^{10}~M_{\odot}$ \citep{natarajan2009there,volonteri2010formation,woods2019titans,latif2016formation,king2015big}. However, the origin of SMBHs becomes more enigmatic with the observations of quasars at $z \gtrsim 6$ \citep{willott2007four,jiang2015discovery}, and the more recent detection of 31 quasars at $6.6 < z < 7.8$ from the Euclid Wide Survey \citep{yang2026euclid}. In particular, \citet{mortlock2011luminous} reported a quasar at $z \sim 7.085$ (0.77 Gyr after the Big Bang) hosting a black hole of $2\times10^9~M_{\odot}$.
	
	These observations imply that the seeds of the first SMBHs must have formed and grown on very short timescales \citep{latif2016formation}. Several formation channels have been proposed to produce SMBH seeds, each with distinct initial masses and associated challenges. The remnants of Population III stars are expected to form light seeds of $\sim 10^2~M_{\odot}$ \citep{fryer2001pair}. These seeds can grow into intermediate-mass black holes (IMBHs) in dense early globular clusters through gas accretion \citep{vesperini2010intermediate}; collisions between Pop III stars can also contribute to building more massive seeds \citep{vergara2021stellar,reinoso2018collisions,reinoso2020effects}, and both mechanisms may occur simultaneously \citep{boekholt2018formation}. However, their growth to SMBH masses requires sustained Eddington accretion or super-Eddington bursts, which are difficult to achieve \citep{regan2024massive,smith2018growth}. The direct collapse of primordial gas clouds can form heavier seeds of $\sim 10^5~M_{\odot}$ \citep{bromm2003formation,lodato2006supermassive,latif2015formation,van2025further}; but this pathway requires specific conditions to suppress fragmentation and maintain effective cooling \citep{shang2010supermassive,suazo2019role}. Moreover, direct collapse black holes (DCBHs) typically form in metal-free environments, and subsequent accretion can be severely suppressed by supernovae (SNe) feedback and by the large relative velocity that the BH acquires due to its formation at $\sim 1~{\rm kpc}$ from the galactic centre \citep{chon2021cosmological}. Runaway collisions in dense stellar clusters provide another pathway, efficiently forming IMBHs with masses of $\sim 10^3$--$10^4~M_{\odot}$, which can serve as seeds for SMBHs \citep{sakurai2017formation,chung2026dense}. However, current simulations of the collisional runaway typically produce IMBHs rather than SMBHs \citep{portegies2002runaway,2023MNRAS.522.4224V,vergara2026efficient,reinoso2018collisions}, partly due to the computational cost of simulating the large number of stars required to model more massive clusters. Once formed, these IMBHs could grow through subsequent gas accretion and mergers with other BHs \citep{inayoshi2022rapid,kritos2023massive,inayoshi2016hyper,chung2026dense,pelupessy2007rapidly}.
	
	Galactic centres are one of the most likely places where SMBHs can form, since the process of mass segregation and dynamical friction (DF) can funnel material to the centre \citep{shlosman1990fuelling,escala2006fueling,naiman2015role}. Besides SMBHs, nuclear stellar clusters (NSCs) can also be found in galactic centres \citep{neumayer2020nuclear,schiavi2021we}, with both referred to under the nomenclature of central massive objects (CMOs; \citet{ferrarese2006fundamental}). There are correlations between properties of the host galaxy and the CMO it contains, including the $M_{\rm BH}$--$\sigma$ relation between BH mass and bulge velocity dispersion \citep{gultekin2009m,tremaine2002slope}, the $M_{\rm BH}$--$M_{\rm bulge}$ relation \citep{haring2004black,marconi2003relation}, as well as correlations between the host galaxy mass and the mass of the SMBH or NSC it harbours \citep{ferrarese2006fundamental}. These correlations suggest that there could be a link in the origin of SMBHs and NSCs \citep{2021ApJ...908...57E}. NSCs dominate in galaxies with masses $\lesssim 10^{10}~M_{\odot}$, while SMBHs dominate in those with masses $\gtrsim 10^{12}~M_{\odot}$; in the intermediate-mass regime, both types coexist \citep{georgiev2016masses}. Runaway collisions are the primary candidate for forming a massive BH in the centre \citep{2021ApJ...908...57E,chung2026dense,10.1093/mnras/stab402}. Mass segregation and the Spitzer instability drive the migration of the most massive stars toward the centre. This increase in the core density triggers a runaway collision process, forming a massive BH as a consequence \citep{gurkan2004formation,rasio2003formation}.
	
	Observations with the James Webb Space Telescope (JWST; \citet{gardner2023james}) have discovered a considerable population of SMBHs at $5 \lesssim z \lesssim 11$ \citep{schneider2023we,natarajan2024first,eilers2023eiger,harikane2025early}, as well as young massive clusters (YMCs) that could eventually form IMBHs through runaway collisions \citep{schleicher2026massive}. The presence of such SMBHs challenges the light seed scenario, since Pop III remnants would require sustained super-Eddington accretion to grow to these masses \citep{kiyuna2026super}. In contrast, heavy seeds can reach SMBH masses through Eddington-limited accretion \citep{jeon2025physical,silk2024came}. Recent observations of extremely red, point-like objects, known as little red dots (LRDs), may help elucidate the origins of early SMBHs, as they could represent an intermediate population between seeds and the observed SMBHs \citep{vaida2026little}. Observations of LRDs span a redshift range of $3 \lesssim z \lesssim 10$. LRDs were initially identified as extremely red galaxies \citep{endsley2023jwst,labbe2023population}. There is no consensus on their nature. Evidence supporting an AGN interpretation includes the detection of high-ionization coronal lines, such as [FeX], and broad H$\alpha$ emission \citep{kocevski2023hidden,furtak2024high}. Although many LRDs are radio and X-ray quiet, this could be expected from super-Eddington accretion models, where thick discs suppress high-energy emission \citep{secunda2026little}. From a stellar interpretation, LRDs are thought to be dusty, intensely star-forming galaxies \citep{escala2025fate,kritos2026nuclear,rantala2026supermassive}. If this is the case, these objects would be among the densest stellar systems ever observed, with core densities as high as $\sim 10^8~M_{\odot}$ pc$^{-3}$, largely exceeding the threshold for runaway collisions, and may include tidal disruption events (TDEs) in some models \citep{bellovary2025little}.
	
	In this paper, we investigate the role of gas in the formation of SMBH seeds. We test the scenario of global instability in NSCs proposed by \citet{2021ApJ...908...57E} through coupled N-body/SPH simulations, by selecting a set of initial conditions from \citet{2023MNRAS.522.4224V} and adding a gas component. We first describe the scenario in Section~\ref{sec:2_coll_scenario} and then present the numerical methods in Section~\ref{sec:3_num_methods}. We present the main results in Section~\ref{sec:results}. We discuss the results and their implications for SMBH seed formation, along with caveats, in Section~\ref{sec:Discussion}. Finally, we summarize the main results and outline considerations for future work in Section~\ref{sec:conclusions}. 
	\section{The collisional runaway scenario for SMBH seeds}
	\label{sec:2_coll_scenario}
	A promising channel for the formation of massive black hole seeds involves runaway stellar collisions in dense stellar clusters \citep{2021ApJ...908...57E}. This scenario is motivated by the observation that NSCs and SMBHs are found in galactic centres \citep{georgiev2016masses}, suggesting a potential evolutionary link. While stellar collisions are negligible in typical galactic fields \citep{2008gady.book.....B}, dense star clusters could have reached densities high enough for the collision timescale ($t_{\rm coll}$) to become shorter than a characteristic timescale $t_{\rm H}$ (e.g., the age of the system or the Hubble time), allowing collisions to become a dominant dynamical process. \citet{2021ApJ...908...57E} established a critical density criterion, derived from $t_{\rm coll} < t_{\rm H}$, to define this unstable, collision-dominated regime. This work showed that observed NSCs generally avoid this parameter space, whereas systems with BHs are consistent with having undergone such a phase.
	\subsection{Timescale criteria and the instability regime}
	The viability of the collisional runaway scenario depends on the interplay between three fundamental timescales: the collision time ($t_{\rm coll}$), the relaxation time ($t_{\rm relax}$), and a characteristic timescale ($t_{\rm H}$). Following the scenario proposed by \citet{2021ApJ...908...57E}, the collisional runaway predicts two global collisional instabilities, both defined by the condition that the collision time is comparable to or shorter than the age of the system ($t_{\rm coll} \leq t_{\rm H}$). In the first regime, defined by systems where $t_{\rm coll} < t_{\rm relax}$ ($\leq t_{\rm H}$), collisions dominate before relaxation can act, leading to runaway collisions that form a massive remnant without the system being able to expand \citep{escala2025fate}. In the second regime, defined by systems where $t_{\rm relax} < t_{\rm coll}$ ($\leq t_{\rm H}$), the system is initially in the unstable regime but expands before collisions are triggered, as two-body relaxation operates on shorter timescales. However, core collapse could be present in this scenario \citep{portegies2002runaway}, enhancing collisions within the core. We now derive the relations that quantify these regimes, following the approach of \citet{2023MNRAS.522.4224V}.
	
	Consider a stellar system of total mass $M$ and characteristic radius $R$, composed of $N$ stars of mass $m$ and radius $r_{\star}$. We assume the system is virialised, with a velocity dispersion $\sigma = (GM/R)^{1/2}$. The collision time can be estimated as $t_{\rm coll} = (n \Sigma \sigma)^{-1}$, where $n = 3M/(4\pi R^3 m)$ is the number density and $\Sigma$ is the effective cross-section \citep{2008gady.book.....B}. Identifying terms in the collisional rate $1/t_{\rm coll} = 16\sqrt{\pi} n \sigma r_{\star}^2(1 + \Theta)$, the cross-section can be defined as $\Sigma_0 = 16 \sqrt{\pi} r_{\star}^2 (1 + \Theta)$, where $\Theta = G m / (2 r_\star \sigma^2)$ is the Safronov number. Substituting these expressions yields $t_{\rm coll} = (4\pi m R^{7/2})/(3 \Sigma_0 G^{1/2} M^{3/2})$.
	
	For a typical velocity dispersion of $\sim 100~{\rm km} {\rm s}^{-1}$, the Safronov number becomes $9.54(m/M_{\odot})(R_{\odot}/r_{\star})$ \citep{2008gady.book.....B}. Consequently, the effective cross-section is determined primarily by the stellar radii, corresponding to the limit where the Safronov number is small. Then, the condition for collisions to be significant over the age of the system, $t_{\rm coll} \leq t_{\rm H}$, leads to a mass-radius relation that serves as the first boundary in an $M$–$R$ diagram:
	\begin{equation}
		M \geq \left(\frac{4 \pi m}{3 \Sigma_0 t_{\rm H} G^{1/2}} \right)^{2/3} R^{7/3} \label{eq:cond_tcoll}.
	\end{equation}
	
	If, instead, we include the virialised velocity dispersion in the Safronov number, the dependence of $\Sigma_0$ on $\sigma$ translates into a dependence on the total mass $M$ through $\sigma=\sqrt{GM/R}$. This distinguishes our derivation from that of \citet{2023MNRAS.522.4224V}, where the cross-section is dominated by stellar radii and the Safronov number is neglected. Moreover, in the absence of an external potential, $M$ denotes the stellar mass. When an external potential of mass $M_{\rm ext}$ is present, the total mass is $M_{\rm tot} = M + M_{\rm ext}$, and the velocity dispersion increases to $\sigma = \sqrt{GM/R}(1+q)$ \citep{reinoso2020effects}, where $q= M_{\rm ext}/ M$. This makes Eq.~\eqref{eq:cond_tcoll} an implicit relation for $M$. Solving this relation for $M$ (see Appendix~\ref{app:Mcrit}) yields:
	\begin{equation}
		\begin{split}
			M_{\rm crit}=& \left(\frac{\beta R^{7/2}}{2} + \sqrt{\frac{\beta^2 R^7}{4} + \frac{\alpha^3 R^3}{27}}\right)^{2/3} \\
			+& \left(\frac{\beta R^{7/2}}{2} - \sqrt{\frac{\beta^2 R^7}{4} + \frac{\alpha^3 R^3}{27}}\right)^{2/3} -   
			\frac{2\alpha R}{3},
		\end{split} 
		\label{eq:cond_tcoll_corr}
	\end{equation}
	where
	\begin{equation}
		\alpha = \frac{9.54 m R_{\odot} v_{100}^2}{r_{\star} G M_{\odot} (1+q)^2},~ \beta = \frac{\sqrt{\pi} m}{12 G^{1/2} r_{\star}^2 (1+q) t_{\rm H}}.
	\end{equation}
	
	Here, $v_{100} = 100~\text{km}~\text{s}^{-1}$. The external potential is assumed to have the same virial radius and mass distribution as the stellar system. Eq. \eqref{eq:cond_tcoll_corr} is the marginal condition for $t_{\rm coll} = t_{\rm H}$. Systems that exceed this mass ($M>M_{\rm crit}$) have $t_{\rm coll} < t_{\rm H}$.
	 
	The relaxation time, which characterizes the timescale for stellar orbits to be significantly altered by cumulative weak encounters, is given by \citet{2008gady.book.....B}: $t_{\rm relax} = t_{\rm cross} \frac{0.1 N}{\ln N}$, where $t_{\rm cross} = R / \sigma$, and $\sigma = \sqrt{GM/R}$ for a virialised system. Replacing the number of stars $N = M/m$, the relaxation time can be written as $t_{\rm relax} =  (0.1 R^{3/2} M^{1/2})/(m G^{1/2}\ln (M/m))$.
	
	The condition $t_{\rm relax} \leq t_{\rm H}$ defines a second relation:
	\begin{equation}
		R \leq \left( \frac{t_{\rm H} m \ln(M/m)}{0.1} \right) ^{2/3} \left( \frac{G}{M} \right) ^{1/3}
		\label{eq:r_relax}.
	\end{equation}
	
	Solving $t_{\rm relax} = t_{\rm H}$ for $M$ including the effects of an external potential (generated by a mass $M_{\rm ext}$) yields the mass scale (see Appendix~\ref{app:Mrelax}):
	\begin{equation}
		M_{\rm relax} = \frac{m}{(1+q)^{2}} \exp\!\left(-2W_{-1}\left(-\frac{1}{2\lambda}\right)\right)~,
		\label{eq:mrelax}
	\end{equation}
	where $W_{-1}$ is the lower branch of the Lambert $W$ function evaluated at $z = -1/(2\lambda)$, and $\lambda = t_{\rm H} \sqrt{mG} / (0.1 R^{3/2}(1+q)^2)$. We use the lower branch of the Lambert function since it is physically more relevant (see Appendix~\ref{ap:solve_Mrelax}). The relation in Eq.~\eqref{eq:mrelax} defines the characteristic mass scale above which a system of radius $R$ can evolve over a time $t_{\rm H}$ without two‑body relaxation processes significantly altering its initial conditions.
	
	The timescales $t_{\rm coll}$ and $t_{\rm relax}$ define three distinct dynamical regimes for a stellar system of mass $M$, radius $R$ and external potential of mass ratio $q=M_{\rm ext}/M$. The unstable regime (UR) occurs when $t_{\rm coll} \leq t_{\rm H} \leq t_{\rm relax}$ (i.e., $M \geq M_{\rm crit}$ and $M \geq M_{\rm relax}$), where collisions dominate before two-body relaxation can expand the cluster, potentially leading to a global collapse and the formation of a massive BH. The stable regime (SR) is defined by $t_{\rm relax} \leq t_{\rm H} \leq t_{\rm coll}$ ($M \leq M_{\rm crit}$ and $M \leq M_{\rm relax}$), where relaxation expands the cluster before collisions become relevant. Finally, we define an intermediate regime (IR) by $t_{\rm relax} \leq t_{\rm coll} \leq t_{\rm H}$ ($M_{\rm crit} \leq M \leq M_{\rm relax}$), where two-body relaxation begins to expand the cluster before global collisions become frequent; however, the core may still undergo collisional collapse while the outer envelope expands. The described regimes characterize the stability of the system against a global collisional collapse. In the UR, we expect that most of the mass ends up in a central BH, resulting in a BH-dominated system. In the SR, the gravitational potential remains dominated by the stellar cluster. The IR represents an intermediate case, where a broad range of BH masses may form embedded in a surrounding NSC.
	\section{Numerical methods and initial conditions}
	\label{sec:3_num_methods}
	We model the coupled evolution of stars and gas using the Astrophysical Multipurpose Software Environment, AMUSE\footnote{https://github.com/amusecode/amuse} \citep{AMUSE_Portegies09, AMUSE_Portegies13, 2013A&A...557A..84P, Portegies2018}. The stellar component is evolved with a pure N-body code, while the gas is modelled with the smoothed particle hydrodynamics (SPH) method \citep{1977MNRAS.181..375G}. The two components interact gravitationally via the \textsc{BRIDGE} method \citep{Fujii2007}. We provide more details below.
	\subsection{Stellar dynamics and collision treatment}
	The stellar dynamics are computed with the pure N-body code \textsc{PH4} \citep{2012ASPC..453..129M}, which implements a 4th-order Hermite integrator \citep{1992PASJ...44..141M}, with adaptive block time steps. The version of \textsc{PH4} used in AMUSE does not include regularization for close encounters (see Section~\ref{sec:Discussion}). Stars are treated as point masses with an associated physical radius.
	
	A collision between two stars is detected when the distance between their centres is less than the sum of their radii, $|\texttt{x}_{i} - \texttt{x}_{j}| < R_i + R_j$. When a collision is detected, the integration is interrupted and the two stars are replaced by a new single star. The merger product inherits the total mass, the centre-of-mass position, and centre-of-mass velocity of the progenitors. The new radius is calculated under the assumption of constant density: $R_{\rm coll} = R_1 ((m_1 + m_2)/m_1)^{1/3}$. This simple treatment conserves linear momentum and mass.
	\subsection{Gas dynamics}
	\label{sec:gas_dynamics}
	The gas is evolved using the SPH code \textsc{FI} \citep{1997A&A...325..972G}, which solves the hydrodynamic equations in Lagrangian form. The code utilizes a cubic spline kernel for interpolation \citep{1992ARA&A..30..543M}, with an adaptive smoothing length to keep a total of 64 neighbors. Self-gravity of the gas is computed using a Barnes-Hut tree code \citep{1986Natur.324..446B}, with a softening length that is equal to the smoothing length. The gas is modelled as an ideal, adiabatic fluid without radiative cooling or external heating.
	
	Although fully conservative equations for entropy and energy are implemented in some SPH codes \citep{springel2002cosmological}, we set up the \textsc{FI} so that it evolves the internal energy, which consequently
	becomes the relevant thermodynamic quantity. For an ideal adiabatic gas with adiabatic index $\gamma$, the equation of state is $P=A \rho^{\gamma}$, where $P$ is the pressure and $\rho$ the density. In SPH, the equation of state is written in terms of the specific internal energy $u$:
	\begin{equation}
		P = (\gamma-1) \rho u~.
	\end{equation}
	
	The sound speed, which is the relevant dynamical quantity in the absence of radiative cooling and heating, is given by \citep{horedt2004polytropes}:
	\begin{equation}
		c_{\rm s} = \sqrt{\frac{\partial P}{\partial \rho}}=\sqrt{\frac{\gamma P}{\rho}} = \sqrt{\gamma(\gamma-1) u}~.
		\label{eq:sound_speed}
	\end{equation}
	\subsection{Gravitational coupling}
	The gravitational interaction between the stellar (\textsc{PH4}) and gaseous (\textsc{FI}) systems is handled by the \textsc{BRIDGE} coupling scheme \citep{Fujii2007}. This method is necessary to efficiently integrate systems with very different intrinsic time scales. \textsc{BRIDGE} uses a symplectic, kick-drift-kick (Verlet) integrator to couple the subsystems. In this scheme, the gravitational force from one subsystem on the other is applied as impulses (kicks) for a half timestep, updating the velocity. Then, each subsystem is evolved independently for a full timestep (drift). Finally, a kick is applied for a half timestep. The gravitational force between gas and stars is calculated with a Plummer softening kernel and with a softening length given by
	\begin{equation}
		\epsilon_{\rm g, s} = 0.5( h + \epsilon_{\rm coup})~,
	\end{equation}
	where $h$ is the smoothing length of the gas particle and $\epsilon_{\rm coup}=5$~AU is the coupling softening length. 
	The initial \textsc{BRIDGE} timestep is chosen to be the crossing time of the stellar component, and we implement an adaptive timestep $\Delta t_{\rm BRIDGE}$. After every \textsc{BRIDGE} timestep, a reference timescale $\tau$ is set to be the minimum between the free-fall time in the densest SPH particle and a Courant-type timestep that depends on the acceleration between gas and star particles (see Appendix~\ref{app:Adaptive_BRIDGE_Timestep}):
	\begin{equation}
		\tau = \min(t_{\rm ff}, \Delta t_{\rm coup, gas}, \Delta t_{\rm coup, stars})~.
	\end{equation}
	
	Finally, the \textsc{BRIDGE} timestep is set to fulfill the condition
	\begin{equation}
		0.5\tau \leq \Delta t_{\rm BRIDGE} \leq \tau~.
	\end{equation}
	
	This adaptive timestep ensures stability against strong gas-stars gravitational interactions.
	\subsection{Initial conditions}
	In order to test the scenario proposed by \citet{2021ApJ...908...57E}, we selected 12 initial conditions for nuclear stellar clusters from \citet{2023MNRAS.522.4224V}, keeping the same model IDs for direct comparison. For each model, we perform four simulations with different gas fractions, using the same stellar initial conditions. Table~\ref{tab:initial_conditions} summarizes the initial conditions for the stellar component, where ID is the model identifier, $R_{v}$ is the initial virial radius, $M$ is the total mass, $N$ is the number of stars, $m$ is the initial star mass, $r$ is the initial star radius and $\sigma$ is the velocity dispersion. Each model is a Plummer distribution \citep{plummer1911problem} with equal-mass stars. The initial virial radii are $R_v=0.005,0.01,0.1$~pc, and the number of stars $N=500, 1000, 5000, 10000$. The stellar masses and radii are the same as in \citet{2023MNRAS.522.4224V}: for $m=1, 10$ and $50~M_{\odot}$, we adopt $r=1, 4.7$ and $11.7~R_{\odot}$, respectively, following the mass–radius relations of \citet{bond1984evolution} and \citet{demircan1991ap}. Due to computational limitations, we simulate clusters that are less massive but more compact than real NSCs. This allows us to explore the instability on shorter timescales ($t_{\rm H} = 5$ Myr) while maintaining the same physical regime. We note that our models are initially significantly denser than observed NSCs (see Appendix~\ref{app:density_comparison}).
	
	Fig.~\ref{fig:SMBH_scenario} shows the initial conditions of models in Table~\ref{tab:initial_conditions} (see Appendix~\ref{app:IC_and_parameters}), in a mass-radius space. The solid lines represent the relaxation mass (Eq.~\eqref{eq:mrelax}), and the dashed lines are for the critical mass (Eq.~\eqref{eq:cond_tcoll_corr}), calculated for the final simulation time ($t_{\rm H}=5$ Myr) for $q=0$; the colored areas indicate the range of gas fractions, from $q=0$ to $q=0.2$. Black dot-dashed line stands for the condition $t_{\rm coll} = t_{\rm relax}$, for $q=0$. Colors indicate the initial star mass, with yellow for $m=1~M_{\odot}$, red for $m=10~M_{\odot}$ and blue for $m=50~M_{\odot}$. Each model is represented by a different symbol, with the same color convention. We note that family models A, B, C, D, H and I lie in the IR (filled symbols), while family models F, J, K, L, M and N are in the SR (open symbols), with no models in the UR.
	\begin{figure}[ht]
		\centering
		\includegraphics[width=0.48\textwidth]{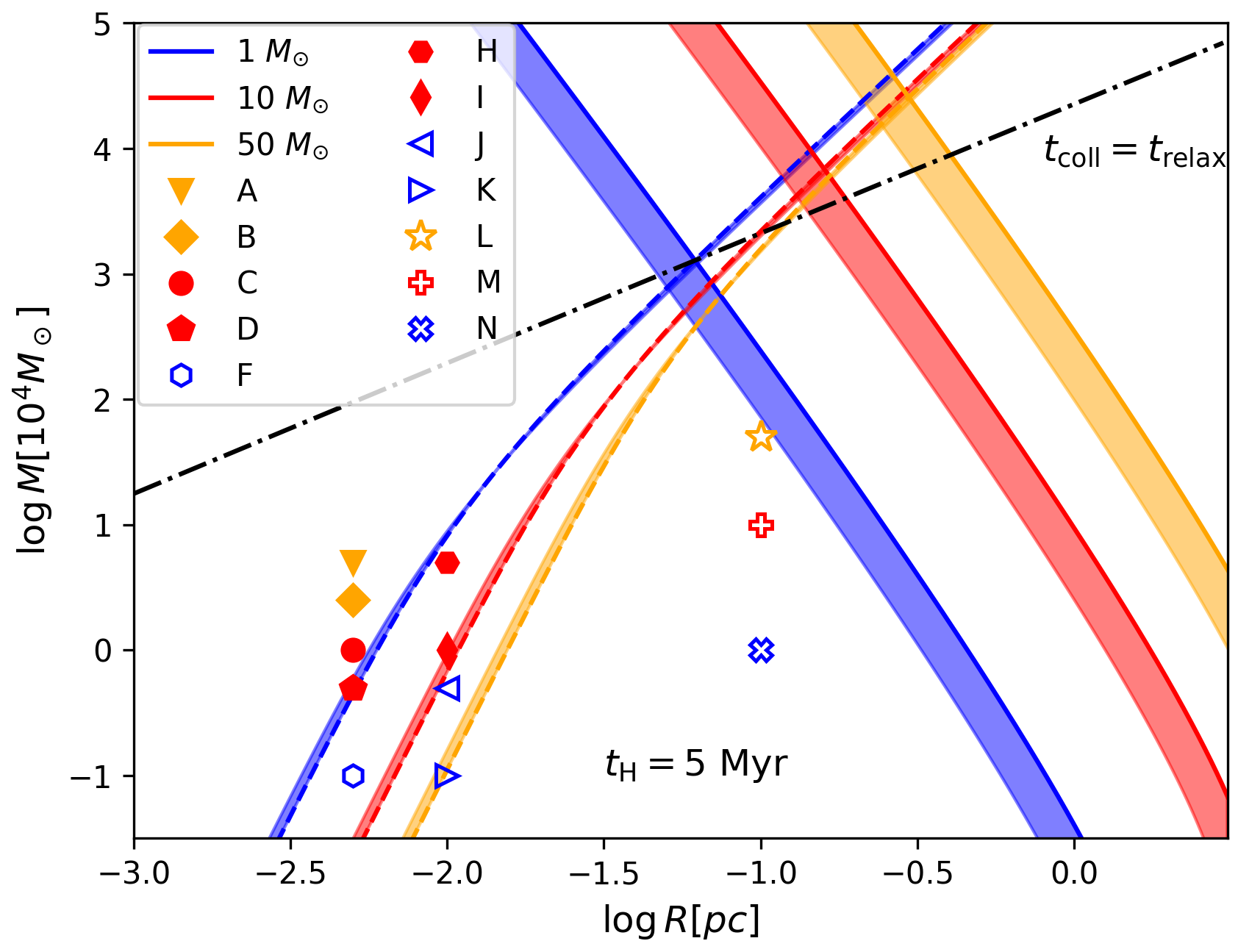}
		\caption{SMBH formation scenario in a mass–radius diagram. Colors indicate the stellar mass: yellow for $m=1~M_{\odot}$, red for $m=10~M_{\odot}$ and blue for $m=50~M_{\odot}$. Solid and dashed lines represent the relaxation and critical masses, respectively, for $t_{\rm H}=5$ Myr and $q=0$. Shaded regions indicate the range of gas fractions, from $q=0$ to $q=0.2$. The black dot-dashed line represents the condition $t_{\rm coll} = t_{\rm relax}$ for $q=0$. Symbols shows the initial conditions from Table~\ref{tab:initial_conditions}: open symbols indicate models in the SR, and filled symbols indicate models in the IR.}
		\label{fig:SMBH_scenario}
	\end{figure}
	
	For each star cluster model, we generate a new Plummer distribution of $N_{\rm SPH}$ SPH particles with the same virial radius and total mass as the stellar system. The number of SPH particles for a given gas fraction $q$ is $N_{\rm SPH}=q \times N \times 100$, with $N$ the number of stars. We then set the SPH particle masses to $m_{\rm SPH}=m/100$, so that the total gas mass becomes a fraction $q$ of the stellar mass, with $q=0.05$, $0.1$ or $0.2$. Contrasting with the stellar component, each gas fraction corresponds to an independent realization, generated from scratch. After this procedure, the gas retains the same virial radius and spatial distribution as the stellar cluster. We note that neither the gas nor the stars are initially in virial equilibrium, since both begin under a potential deeper than the equilibrium one, with total mass $M(1+q)$. We show in Section~\ref{sec:results} that this imbalance is transient and does not drive the subsequent evolution.
	
	To characterize the gas further, we consider the initial sound speed profiles (Eq.~\eqref{eq:sound_speed}) for the gas variants of the A family (A$_{5}$, A$_{10}$ and A$_{20}$) and the N family (N$_{5}$, N$_{10}$ and N$_{20}$), which are the most and least collisional models families in our sample (in their respective regimes), respectively. For the A gas variants, the stellar velocity dispersion decreases from $186.80~{\rm km}~{\rm s}^{-1}$ in the central region ($r_{5}$) to $101.60~{\rm km}~{\rm s}^{-1}$ in the outer regions ($r_{95}$), while the gas sound speed falls from $129.89~{\rm km}~{\rm s}^{-1}$ at the centre to $65.51~{\rm km}~{\rm s}^{-1}$ in the outermost layers (see Appendix~\ref{app:extra_plots}).
	
	The fact that $\sigma/c_{\rm s}>1$ throughout the stellar cluster has important consequences. The transonic motion of stars through the gas produces an efficient DF, as described by \citet{ostriker1999dynamical}, which leads to stars losing orbital energy and sinking toward the centre, enhancing the occurrence of collisions in the central region. Although our simulations do not fully resolve DF (see Appendix~\ref{app:dynamical_friction}), the dissipation of kinetic energy due to gas-star interactions is still present. For the L gas variants (L$_{5}$, L$_{10}$ and L$_{20}$), $\sigma$ and $c_{\rm s}$ lie in the range $\sim 50$--$125~{\rm km}~{\rm s}^{-1}$, and the condition $\sigma/c_{\rm s}>1$ still holds.
	\section{Results}
	\label{sec:results}
	Here we present the main results from our simulations. We use the black hole efficiency defined in \citet{2021ApJ...908...57E}:
	\begin{equation}
		\varepsilon_{\rm BH} = (1 + M_{\rm NSC}/M_{\rm BH})^{-1}~,
		\label{eq:eff_bh}
	\end{equation}
	where $M_{\rm NSC}$ is the mass of gravitationally bound stars and $M_{\rm BH}$ is the black hole (BH) mass. Since our simulations do not include collapse of stars into BHs, $\varepsilon_{\rm BH}$ is computed using $M_{\rm MMO}$ as the BH mass (see Section~\ref{Disc:caveats}). Considering $M_{\rm Esc}$ as the escaped stars mass, the total mass of the stellar system is $M = M_{\rm Esc} + M_{\rm NSC} + M_{\rm BH}$, and the gravitationally bound mass corresponds to $M_{\rm NSC} + M_{\rm BH}$. We can note that at the beginning of the simulation, $M = M_{\rm NSC}$. In terms of the efficiency, the BH candidate mass is then $M_{\rm BH} = \varepsilon_{\rm BH} (M - M_{\rm Esc})$.
	\subsection{Dynamical evolution}
	In this section we present the time evolution of the mass components of the system. We select two model families from different regimes, A and L. For each family, we consider two variants: a stars‑only simulation (denoted A$_{0}$ and L$_{0}$), and simulations with gas masses of 20\% of the initial stellar mass (denoted A$_{20}$, L$_{20}$). The A variants have an initial stellar mass of $M=5 \times 10^4~M_{\odot}$, virial radius $R_v = 0.005$ pc, and is composed of $10^3$ stars. The L variants have an initial mass of $M=5 \times 10^5~M_{\odot}$, virial radius $R_v = 0.1$ pc and $10^4$ stars. As shown in Fig.~\ref{fig:SMBH_scenario}, the A family is in the IR, where the collision time is shorter than the simulation time. In contrast, the L family is in the SR where collisions are not globally relevant.
	
\begin{figure}[ht]
		\centering
		\includegraphics[width=0.5\textwidth]{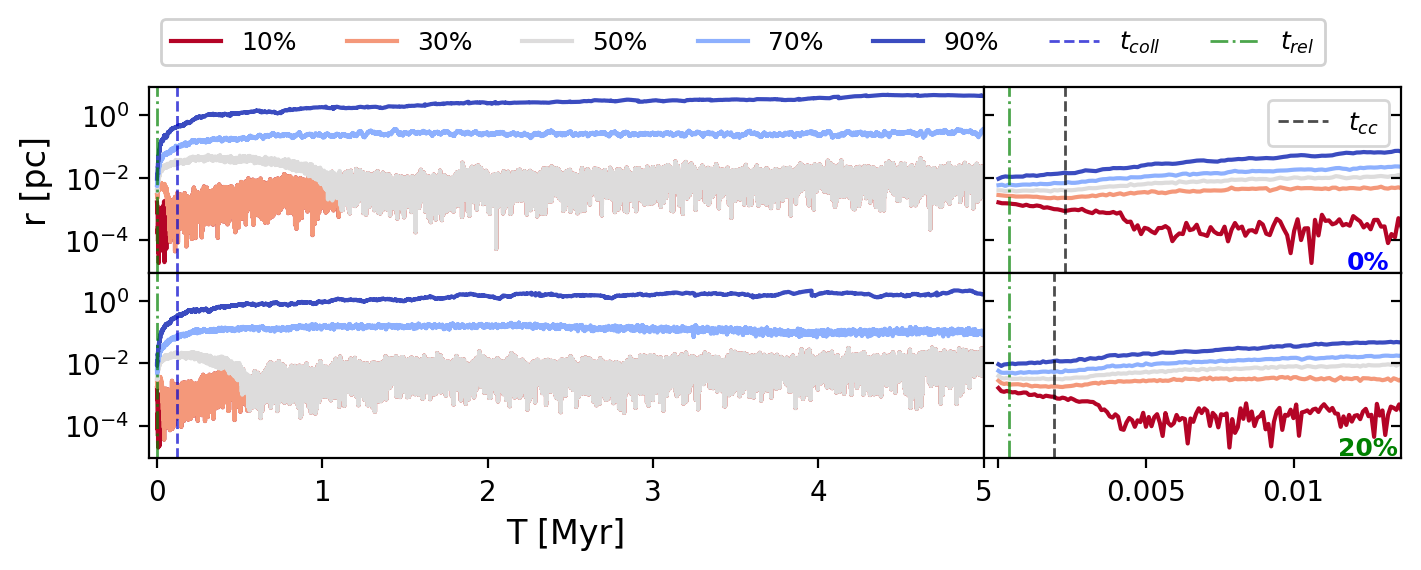}
		\caption{Lagrangian radii for stars, from 10\% (red) to 90\% (blue). The top panel corresponds to A$_{0}$, and the bottom panel to A$_{20}$. The left column shows the full simulation time (5 Myr), while the right column shows the evolution up to $40~t_{\rm relax}$. Blue dashed and green dot-dashed lines are the initial collision and relaxation timescales, respectively. Black dot-dashed lines are the core collapse.}
		\label{fig:Lrad_A_stars}
	\end{figure}

	In Fig.~\ref{fig:Lrad_A_stars}, the top panel corresponds to A$_0$ and the bottom panel to A$_{20}$. The left column shows the evolution over the full 5 Myr, while the right column zooms in to 40 relaxation timescales. The dashed blue line marks the initial collision time, the dot-dashed green line indicates the initial relaxation time, and the dashed black line shows the estimated core collapse time (see Appendix~\ref{app:cc_detection}). We observe that the inner 10\% and 30\% Lagrangian radii rapidly contract towards the centre, on scales of the collision time. After 1 Myr, A$_0$ and A$_{20}$ show that the Lagrangian radii remain stable, with A$_0$ showing a slight increase in the 90\% Lagrangian radius relative to A$_{20}$. In the right column we can observe slight differences in the core collapse (slightly shorter for A$_{20}$), and no considerable changes during 1 $t_{\rm relax}$. This indicates that the initial subvirial perturbation of the stellar component for A$_{20}$, which acts on a dynamical timescale ($< t_{\rm relax}$), does not significantly affect the subsequent evolution.
	
\begin{figure}[t]
		\centering
		\includegraphics[width=0.5\textwidth]{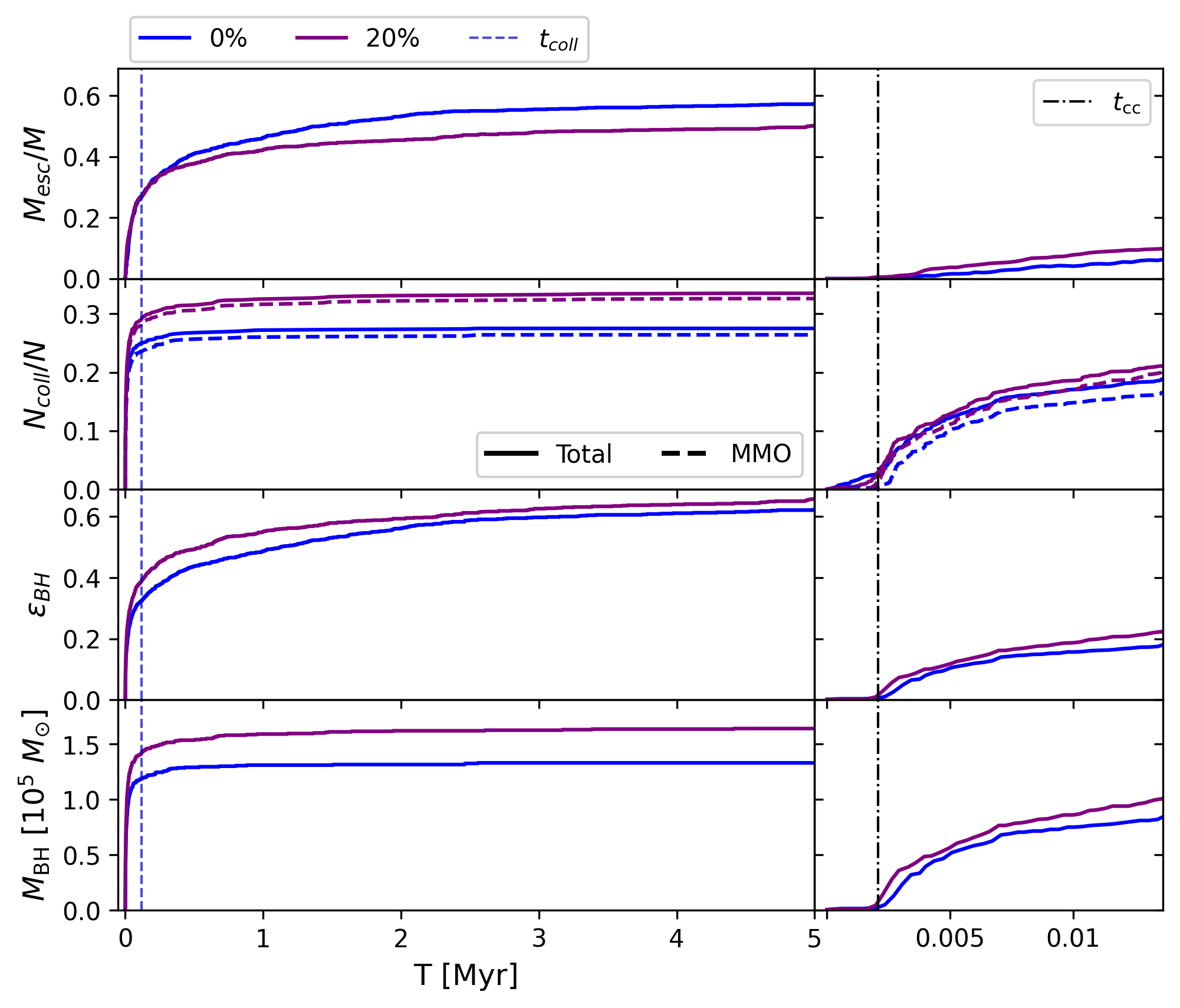}
		\caption{Time evolution of the mass components and $\varepsilon_{\rm BH}$ for stars. Blue curves are for A$_0$, and purple for A$_{20}$. The top panel shows the escaped mass normalised by the total initial stellar mass. The second panel shows collisions normalised by the initial number of stars, where solid lines are for the total and dashed lines for collisions involving the most massive object (MMO). The third panel shows $\varepsilon_{\rm BH}$ (Eq.~\ref{eq:eff_bh}), and the bottom panel shows the mass of the MMO. The left column shows the full 5 Myr, with the vertical blue dashed line representing $t_{\rm coll}$. The right column shows a zoom on $40~t_{\rm relax}$ ($\sim 0.014$ Myr), where the dot-dashed black line marks the mean core collapse time, averaged over the four gas variants of model A.}
		\label{fig:Mass_evol_A}
	\end{figure}

	The time evolution of the mass components and $\varepsilon_{\rm BH}$ of stars for A$_{0}$ and A$_{20}$ is displayed in Fig.~\ref{fig:Mass_evol_A}. The top panel shows the escaped mass normalised by the initial stellar mass $M$. We find that $\sim 45\%$ of the initial stellar mass escapes during the first Myr of evolution for both variants. By 5 Myr, the A$_{0}$ variant shows 57.3\% of escaped mass, compared to 50.2\% for A$_{20}$. The second panel shows collisions normalised by the initial number of stars, with solid lines for total collisions and dashed lines for the cumulative number of collisions in the progenitor merger tree of the MMO. We can see that most collisions occur within $t_{\rm coll}$ (vertical dashed blue line), increasing as the gas mass increases. During the first Myr, A$_{0}$ shows 27.3\% of total collisions, and 27.6\% at 5 Myr. A$_{20}$ shows 32.6\% of collisions at 1 Myr, and 33.6\% of collisions at 5 Myr. The influence of gas is stronger during the early evolution of the cluster, increasing the number of collisions by 19.41\% relative to A$_{0}$. The gas also increases the number of collisions at late times. The third panel shows the evolution of $\varepsilon_{\rm BH}$ (Eq.~\eqref{eq:eff_bh}), which is 62.30\% and 65.86\% for A$_{0}$ and A$_{20}$, respectively. As described in \citet{2023MNRAS.522.4224V}, the increase in efficiency in the late stage of the system is dominated by the escaped mass, as collisions are less common. The bottom panel shows the mass evolution of the MMO (our BH candidate), exhibiting a similar behaviour to the collision panel, with a rapid increase at early times, and a negligible increase in late evolution. The final mass of the MMO is 13\,300 $M_{\odot}$ for A$_{0}$, and 16\,400 $M_{\odot}$ for A$_{20}$, representing 26.6 and 32.8\%, respectively, of the initial stellar mass, which means that most collisions end in the MMO. The right column is a zoom at $40~t_{\rm relax}$, where the black dash-dotted line marks the mean core collapse time, averaged over the four gas variants, which coincides with the onset of runaway collisions. In the second panel, the total number of collisions (solid) rises faster than the cumulative number of collisions in the progenitor merger tree of the MMO (dashed). This difference reflects collisions occurring elsewhere in the cluster at early times, which build up small objects. These objects subsequently merge with the MMO, so that by $40~t_{\rm relax}$ almost all collisions involve the MMO. 
	
\begin{figure}[ht]
		\centering
		\includegraphics[width=0.5\textwidth]{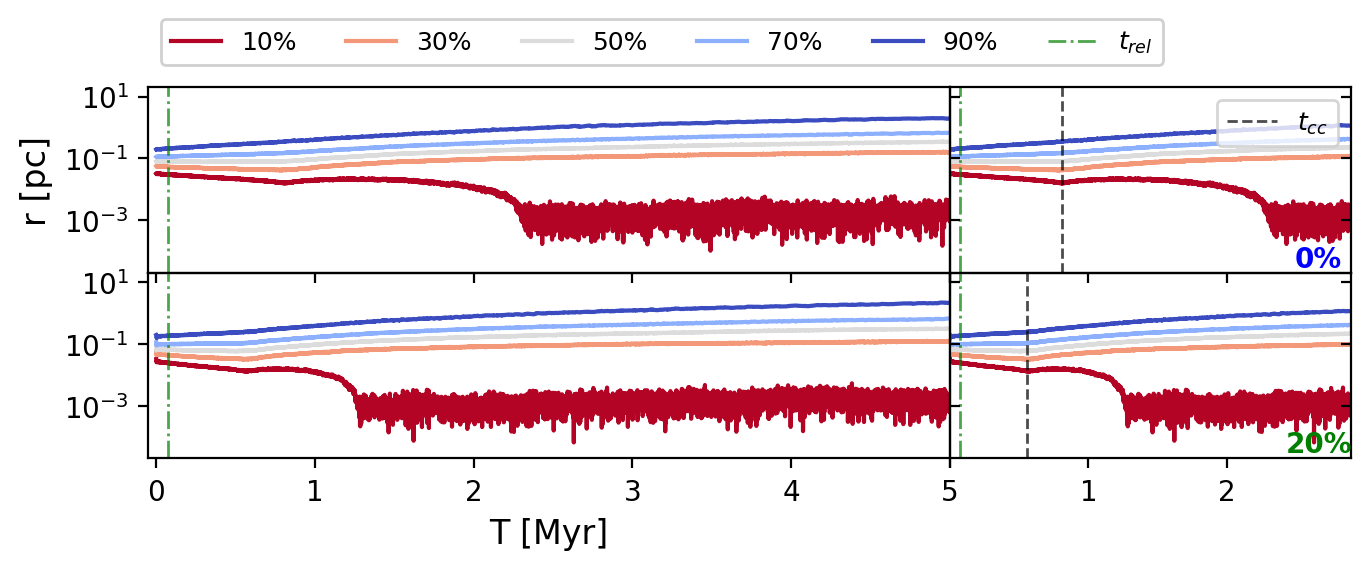}
		\caption{Lagrangian radii for stars, from 10\% (red) to 90\% (blue). The top panel corresponds to L$_{0}$, and bottom panel to L$_{20}$. Green and black dot-dashed line are the initial relaxation timescales and core collapse, respectively.}
		\label{fig:Lrad_L_stars}
	\end{figure}

	Fig.~\ref{fig:Lrad_L_stars} shows the Lagrangian radii of L$_0$ and L$_{20}$. For L$_{0}$, the 90\% and 70\% radii expand slowly, while the 30\% and 10\% radii contract slightly. After core collapse, only the 10\% Lagrangian radius continues to contract. L$_{20}$ exhibits qualitatively the same behaviour, with a faster core collapse than L$_{0}$.
	
\begin{figure}[ht]
		\centering
		\includegraphics[width=0.5\textwidth]{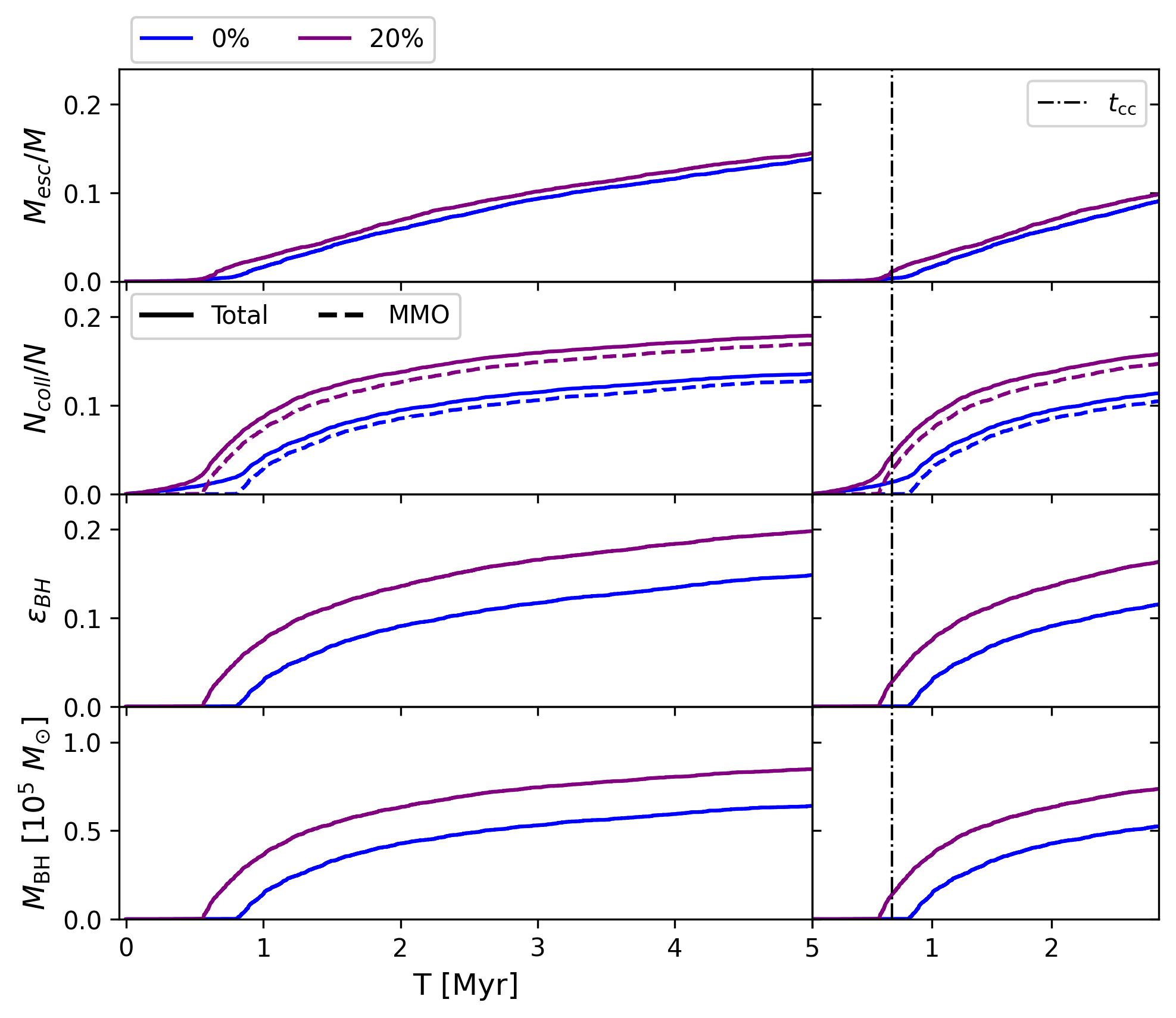}
		\caption{Mass components evolution over time of stars for four variants of model L. Equivalent to Fig.~\ref{fig:Mass_evol_A} with no zoom.}
		\label{fig:Mass_evol_L}
	\end{figure}

	The mass and $\varepsilon_{\rm BH}$ evolution of L$_0$ and L$_{20}$ is presented in Fig.~\ref{fig:Mass_evol_L}. The top panel shows that at 5 Myr the escaped mass is $\sim 14\%$. From the second panel, the normalised number of collisions in the MMO is $\sim 12.79\%$ for L$_{0}$ and $\sim 16.93\%$ for L$_{20}$. Consequently, the BH formation efficiencies are $\sim 14.86\%$ and $\sim 19.74\%$, respectively. Since the escaped mass is similar for both L$_{0}$ and L$_{20}$, the difference in $\varepsilon_{\rm BH}$ reflects the more massive BH candidate formed in L$_{20}$. Specifically, the mass of the MMO is $\sim 6.4\times 10^4~M_{\odot}$ for L$_{0}$ and $\sim 8.5\times 10^4~M_{\odot}$ for L$_{20}$.
	
	Family L evolves on a much slower timescale than family A: $40~t_{\rm relax}$ corresponds to $\sim 57.91\%$ of 5 Myr simulation for family L, and $\sim 0.27\%$ for family A. Comparing both families over this fixed number of relaxation times therefore ensures that they are at equivalent stages of their dynamical evolution, despite the differences in timescales. At $40~t_{\rm relax}$, the escaped mass is comparable across all four variants, ranging from $\sim6\%$ (A$_{0}$) to $\sim10\%$ (A$_{20}$, L$_{0}$, L$_{20}$). The cumulative collisions in the MMO are lower for family L ($\sim 10$--$15\%$) than for family A ($\sim17$--$20\%$). Consequently, family A shows higher efficiencies ($\sim17$--$22\%$) than family L ($\sim11$--$16\%$).
	
	\subsection{Effect of gas on core collapse timescale}
	For each model, we estimated the core collapse time $t_{\rm cc}$ using the method described in Appendix~\ref{app:cc_detection}. Table~\ref{tab:core_collapse} shows the ratio $x_{\rm cc} = t_{\rm cc, gas}/t_{\rm cc,0}$ averaged over models, where the columns, from left to right are for models in the IR, SR and all, respectively. Each row corresponds to a gas fraction. In the case of models in the IR, for a gas fraction of 5\%, the mean ratio is $0.8765 \pm 0.1818$; for 10\%, $0.8345 \pm 0.1345$; and for 20\%, $0.7410 \pm 0.1094$. All ratios are below 1, indicating a reduced $t_{\rm cc}$ in the presence of gas. The core collapse time tends to be shorter as the gas fraction increases, while the dispersion decreases. The decrease in dispersion with higher gas fraction suggests that the effect becomes more predictable when more gas is present.
	\begin{table*}[h]
		\centering
		\caption{Mean ratio $x_{\rm cc}=t_{\rm cc, gas}/t_{\rm cc, 0}$ of core collapse times for gas-embedded models relative to their gas-free counterparts, and standard deviation for IR family models (A, B, C, D, H, I), SR family models (F, J, K, L, M), and the combined sample.}
		\label{tab:core_collapse}
		\begin{tabular}{c c c c c c c}
			\toprule
			& \multicolumn{2}{c}{IR} & \multicolumn{2}{c}{SR} & \multicolumn{2}{c}{Total} \\
			\cmidrule(lr){2-3} \cmidrule(lr){4-5} \cmidrule(lr){6-7}
			$q$ & $\overline{x}_{\rm cc}$ & $\sigma$ & $\overline{x}_{\rm cc}$ & $\sigma$ & $\overline{x}_{\rm cc}$ & $\sigma$ \\
			\midrule
			0.05  & 0.8765 & 0.1818 & 1.0111 & 0.1570 & 0.9377 & 0.1836 \\
			0.1 & 0.8345 & 0.1345 & 0.8571 & 0.1068 & 0.8448 & 0.1232 \\
			0.2 & 0.7410 & 0.1094 & 0.7082 & 0.0533 & 0.7261 & 0.0899 \\
			\bottomrule
		\end{tabular}
		\begin{tablenotes}
			\item N is excluded from the SR average because it does not reach core collapse within the 5~Myr simulation time.
		\end{tablenotes}
	\end{table*}
	
	For models in the SR, the mean ratio is $1.0111 \pm 0.1570$ for 5\% gas fraction, $0.8571 \pm 0.1068$ for 10\% gas fraction, and $0.7082 \pm 0.0533$ for 20\% gas fraction. The reduction of $t_{\rm cc}$ is less pronounced for SR models at 5\% and 10\% gas fractions, and at 5\% it is almost negligible. The dispersions are smaller for all gas fractions compared with models in the IR. For a gas fraction of 20\%, the reduction in $t_{\rm cc}$ is slightly more pronounced in the SR models ($0.7082 \pm 0.0533$) than in the IR models ($0.7410 \pm 0.1094$). Despite these small differences between IR and SR models, the general tendency is to shorten the core collapse time as the gas mass increases. 
	
\begin{figure}[ht]
		\centering
		\includegraphics[width=0.5\textwidth]{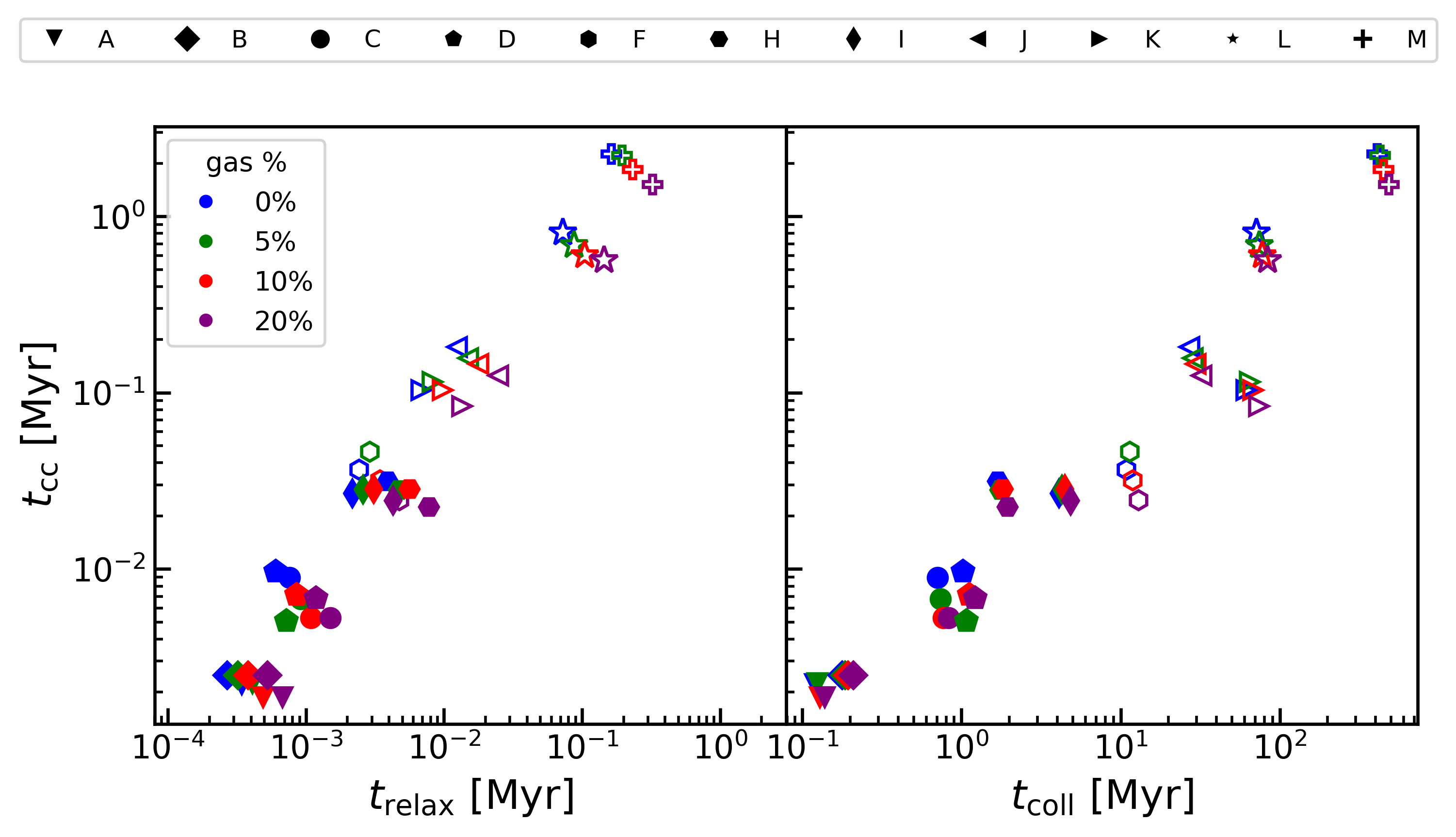}
		\caption{Core collapse time as a function of $t_{\rm relax}$ (left panel) and $t_{\rm coll}$ (right panel).}
		\label{fig:tcc_vs_trel_tcoll}
	\end{figure}

	Fig.~\ref{fig:tcc_vs_trel_tcoll} shows the core collapse time as a function of the relaxation time and collision time, on left and right panels respectively. In addition to the described behaviour, we observe correlations between $t_{\rm cc}$ and both the relaxation and collision times: $t_{\rm cc}$ decreases with shorter $t_{\rm relax}$ (left panel) and with shorter $t_{\rm coll}$ (right panel).
	\subsection{Critical mass and BH formation efficiency}
	In this subsection, we analyse the black hole formation efficiency of our models. First, we compare our stars‑only results with those of \citet{2023MNRAS.522.4224V} to validate our setup. Then, we show how the presence of gas modifies the efficiency as a function of the critical mass, testing the scenario proposed by \citet{2021ApJ...908...57E}.
	
	To make a first comparison, Table~\ref{tab:comparison_vergara} presents results for the same models as \citet{2023MNRAS.522.4224V}. From left to right, the columns are: model identifier, escaped mass (in $M_{\odot}$), black hole mass (in $M_{\odot}$) and formation efficiency ($\varepsilon_{BH}$). For each quantity, two sub‑columns show our stars‑only results (left) and those of \citet{2023MNRAS.522.4224V} (right).
	\begin{table*}[h]
		\centering
		\normalsize
		\caption{Comparison of stars‑only (sub zero variants) simulations with \citet{2023MNRAS.522.4224V}. Our values are from single realizations; reference values include standard deviations from three realizations.}
		\label{tab:comparison_vergara}
		\begin{tabular}{l 
				S[table-format=5.0] 
				S[table-format=5.0(4)] 
				S[table-format=5.0] 
				S[table-format=5.0(4)] 
				S[table-format=2.2] 
				r}
			\toprule
			Model ID & \multicolumn{2}{c}{$M_{\rm Esc}$ [$M_{\odot}$]} & \multicolumn{2}{c}{$M_{\rm BH}$ [$M_{\odot}$]} & \multicolumn{2}{c}{$\varepsilon_{\rm BH}$ [\%]} \\
			\cmidrule(lr){2-3} \cmidrule(lr){4-5} \cmidrule(lr){6-7}
			& {This work} & {Vergara+23} & {This work} & {Vergara+23} & {This work} & {Vergara+23} \\
			\midrule
			A$_0$  & 28650 & 28533(147) & 13300 & 9917(143) & 62.30 & 46.33 $\pm$ 2.5\\
			B$_0$  & 13300 & 14300(687) & 5250 & 3850(324) & 44.87 & 36.00 $\pm$ 3.7\\
			C$_0$  & 5620  & 5370(42)   & 1920 & 1660(14)  & 43.84 & 35.67 $\pm$ 0.5\\
			D$_0$  & 2540  & 2767(111)  & 630  & 610(43)   & 25.61 & 27.33 $\pm$ 3.1\\
			F$_0$  & 411   & 372(5)     & 88   & 85(12)    & 14.94 & 13.67 $\pm$ 2.1\\
			H$_0$  & 21750 & 23137(1966) & 12100 & 11397(143) & 42.83 & 42.67 $\pm$ 2.1\\
			I$_0$  & 4500  & 4253(183)  & 980  & 1110(83)  & 17.82 & 19.33 $\pm$ 1.7\\
			J$_0$  & 1469  & 1459(22)   & 435  & 545(42)   & 12.32 & 15.67 $\pm$ 1.2\\
			K$_0$  & 344   & 327(8)     & 59   & 59(2)     & 8.99  & 8.67 $\pm$ 0.5\\
			L$_0$  & 69300 & 75517(1812) & 64000 & 67450(2585) & 14.86 & 15.67 $\pm$ 0.5\\
			M$_0$  & 7170  & 7850(283)  & 5510 & 6987(246) & 5.94  & 7.67 $\pm$ 0.5\\
			N$_0$  & 41    & 234(44)    & 3    & 71(5)     & 0.03  & 0.73 $\pm$ 0.0\\
			\bottomrule
		\end{tabular}
		\begin{tablenotes}
			\item This work: simulations run up to 5~Myr. Vergara+23: simulations of \citet{2023MNRAS.522.4224V}, run up to 10~Myr. The stellar initial conditions for each of our models correspond to the same letter in Vergara+23; e.g., our model A$_0$ shares the same stellar initial conditions as model A in Vergara+23.
		\end{tablenotes}
	\end{table*}
	For the escaped mass, the differences are small: the ratio of our values to those of \citet{2023MNRAS.522.4224V} ranges from less than 1\% to approximately 10\%, corresponding to less than 5\% of the initial stellar mass. In the case of the BH mass, the largest differences occur for the most collisional models (A$_0$, B$_0$, C$_0$, in the IR), with relative differences of 34.11\%, 36.36\%, and 15.66\%, respectively. These correspond to differences of 6.77\%, 5.60\%, and 2.60\% of the initial stellar mass.
	
	These differences probably arise from the different numerical codes used. Our simulations employ \textsc{PH4} (via AMUSE), while \citet{2023MNRAS.522.4224V} use NBODY6++GPU \citep{wang2015nbody6++,aarseth2003gravitational}. Both use a fourth‑order Hermite integrator with hierarchical block time‑steps. However, NBODY6++GPU additionally implements the Kustaanheimo–Stiefel algorithm \citep{kustaanheimo1965perturbation} and chain regularization \citep{mikkola1993implementation} as special treatments for close encounters, binaries, and multiple systems.
	
	The impact on BH mass translates into efficiency differences. Again, the most collisional models (A$_0$, B$_0$, C$_0$) show a considerable enhancement in BH formation efficiency, with differences of 15.97, 8.87, and 8.17 percentage points, respectively. For the remaining models, the differences are below 4\% percentage points.
	
	Significant differences are observed for N$_0$. In our simulations, our values are lower by 82.48\% and 95.77\% for escaped mass and BH mass, respectively. This discrepancy arises because the relaxation time of model N is relatively long ($\sim 0.5$ Myr) and our simulation time is limited to 5 Myr. In contrast, models L$_0$ and M$_0$, despite also having large relaxation times, reach core collapse within 5 Myr (at $\sim 0.9$ and $\sim 2.3$ Myr, respectively).
	
\begin{figure*}[h]
		\centering
		\includegraphics[width=0.7\textwidth]{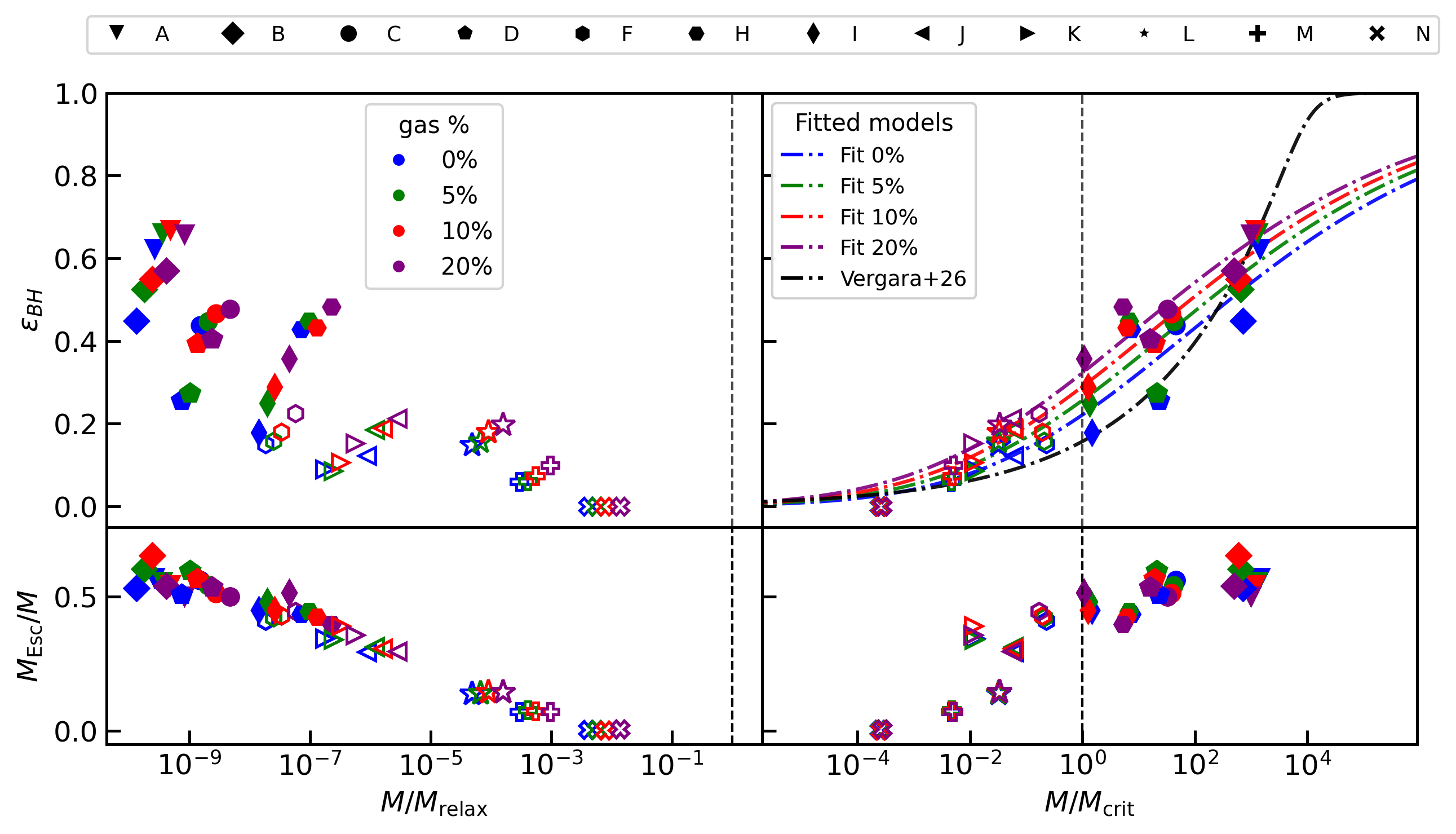}
		\caption{In the y-axis, top panel shows the BH efficiency ($\varepsilon_{\rm BH}$), and bottom panel the normalised escaped mass ($M_{\rm Esc}/M$). The left column shows quantities as a function of $M/M_{\rm relax}$, while the right column shows quantities as a function of $M/M_{\rm crit}$. Symbols correspond to the different models (see legend above the figure), and colors indicate the gas fraction: blue (0\%), green (5\%), red (10\%), purple (20\%). Dashed black lines mark mass ratios equal to unity. The dash-dotted curves in the top-right panel are fits of an asymmetric sigmoid function (Eq.~\eqref{eq:ebh_model}) for each gas fraction, sharing the same shape parameters $x_0$, $k$, $a$ (fitted to the stars‑only data) while $\alpha_q$ is adjusted for each gas fraction. The black dash-dotted curve is the model from \citet{vergara2026efficient}.}
		\label{fig:eff_mesc_Mrate}
	\end{figure*}

	Next, we show results for all models, including their gas variants. Fig.~\ref{fig:eff_mesc_Mrate} shows $\varepsilon_{\rm BH}$ (top panel), and normalised escaped mass (bottom panel). In the left column, quantities are plotted as a function of $M/M_{\rm relax}$, while in the right column they are plotted as a function of $M/M_{\rm crit}$. From the bottom panels, we observe a tendency for the escaped mass to grow as $M/M_{\rm relax}$ decreases and $M/M_{\rm crit}$ increases, with no clear distinction between gas variants. There are two processes that produce escaping stars: diffusive or small-angle relaxation evaporates stars slowly from the cluster; on the other hand, two-body encounters produce ejections of high-energy stars, more likely in dense cores \citep{khalisi2007comprehensive}. Family models in the IR that are more collisional (A, B, and C) have shorter core collapse and collision times, showing an important fraction of escapers (probably due to ejections). In the other extreme, the N family models, where core collapse is not reached and collisions are not globally relevant, exhibit a small fraction of escapers.
	
	In the top-right panel, we observe a clear tendency for efficiency to grow as $M/M_{\rm crit}$ increases. All models follow the same overall trend, but with a systematic vertical shift: the stars‑only variants (blue) lie systematically below the fitted curves, while the 20\% gas variants lie above. This is confirmed by the fits. Near the first regime ($M/M_{\rm crit} < 10^{-4}$) and the third regime ($M/M_{\rm crit} > 10^{5}$), where the efficiency approaches 0 and 1 respectively, these differences become smaller, as all curves converge. To model this effect, we performed a unified fit where the shape of the curve is determined by a common set of parameters ($x_0$, $k$, $a$), while the gas fraction $q$ introduces a horizontal scaling of the mass ratio. We define an effective mass ratio $x_{\rm eff} = (M/M_{\rm crit})(1+\alpha_q q)$, where $\alpha_q$ is a free parameter that depends on $q$. Similar to \citet{vergara2026efficient}, the efficiency is then modelled by the form:
	\begin{equation}
		\varepsilon_{\rm BH} = f(x_{\rm eff}) = \left(1 + \exp\left[-k \left(\log_{10}(x_{\rm eff}) - \log_{10}(|x_0|)\right)\right]\right)^{a}~.
		\label{eq:ebh_model}
	\end{equation}
	
	The parameters $x_0$, $k$ and $a$ were obtained by fitting the stars-only variants ($q=0$). For each gas fraction, we then fitted a single parameter $\alpha_q$. The resulting parameters are $x_0 = 0.039$, $k=0.349$, $a=-3.163$ and $\alpha_q=26.7,~35.6,~47.8$ for $q = 0.05,~0.1,~0.2$, respectively. The fitted function shows three regimes. For $M/M_{\rm crit} < 10^{-4}$, the efficiency is effectively 0. Between $10^{-4}$ and $10^5$, the efficiency increases approximately linearly with $\log_{10}(M/M_{\rm crit})$, reaching 70-80\%. For $M/M_{\rm crit} > 10^{5}$, it asymptotically approaches unity. The main differences appear in the intermediate regime, where the efficiency increases with gas fraction. Compared with the fitted model of \citet{vergara2026efficient}, our models exhibit a smoother transition between the regimes described above.
	\begin{table*}[t]
		\centering
		\caption{Ratios of gas runs to stars-only runs. Values are means $\pm$ standard deviation over all models.}
		\label{tab:gas_ratios}
		\begin{tabular}{c 
				S[table-format=1.4(4)] 
				S[table-format=1.4(4)] 
				S[table-format=1.4(4)]}
			\toprule
			Gas fraction & {$\varepsilon_{\rm BH, gas}/\varepsilon_{\rm BH, 0}$} & {$M_{\rm BH}/M_{\rm BH,0}$} & {$M_{\rm Esc}/M_{\rm Esc,0}$} \\
			\midrule
			5\%  & 1.1256(0.1627) & 1.0840(0.1589) & 1.0471(0.0633) \\
			10\% & 1.2701(0.2079) & 1.2251(0.2057) & 1.0442(0.0905) \\
			20\% & 1.4510(0.3139) & 1.4414(0.2132) & 1.0042(0.0869) \\
			\bottomrule
		\end{tabular}
	\end{table*}
	
	Table~\ref{tab:gas_ratios} lists the mean values for the normalised escaped mass and efficiencies, where each gas variant is normalised by its corresponding stars-only counterpart. The results are consistent with Fig.~\ref{fig:eff_mesc_Mrate}. For the escaped mass, the means are within one standard deviation of unity for all gas fractions, indicating no significant difference. In particular, for the 20\% gas case, the mean ratio is closest to 1 and the dispersion is the smallest among all gas fractions. The black hole mass ratios show a clearer trend. The mean ratios are 1.0840, 1.2251 and 1.4414 for 5\%, 10\%, and 20\% gas, corresponding to increases of 8.40\%, 22.51\%, and 44.14\% relative to the stars-only cases. Since the variance in escaped mass is negligible compared to the increases in BH mass, the efficiency ratios follow a similar trend, with mean values of 1.1256, 1.2701, and 1.4510 (increases of 12.56\%, 27.01\%, and 45.10\%) for 5\%, 10\%, and 20\% gas, respectively.
	\section{Discussion}
	\label{sec:Discussion}
	In this work, we have performed simulations for a set of initial conditions that include coupled star-gas variants. We studied the effect of gas on SMBH seed formation in nuclear stellar clusters. Using the Astrophysical Multipurpose Software Environment (AMUSE), we coupled an N-body integrator (\textsc{PH4}) with an SPH code (\textsc{FI}) via the \textsc{BRIDGE} scheme. In this implementation, the interaction between gas and stars is purely gravitational. Using a selection of initial conditions from \citet{2023MNRAS.522.4224V} with different gas fractions, we tested the scenario proposed by \citet{2021ApJ...908...57E}. Although our stars-only simulations show some differences with respect to \citet{2023MNRAS.522.4224V}, the simulations with gas show an enhancement in black hole mass and efficiency, as well as a shortening of core collapse.
	\subsection{The role of gas on SMBH formation}
	\label{sec:role_of_gas}
	Stellar collisions in dense star clusters have been studied extensively in N-body simulations. \citet{portegies1999star} simulated young star clusters and showed that collisions are more frequent than expected from simple timescale estimates, and that they rejuvenate stars, extending their lifetimes before supernova explosions. As stars interact, massive stars tend to reach the core due to mass segregation over the half-mass relaxation timescale \citep{1987degc.book.....S}. Mass segregation also implies that massive objects, such as binaries or massive stars, can return to the core after being ejected by strong encounters, as a result of DF \citep{1987degc.book.....S,hut1992binaries}. The products of stellar collisions lose kinetic energy, causing them to sink toward the centre. DF and mass segregation explain why most collisions involve a single object, and therefore make this an effective mechanism for BH formation. Numerical simulations have shown that energy equipartition is not reached under realistic initial conditions \citep{trenti2013no}. This implies that massive stars remain dynamically decoupled from the lighter population, continuing to segregate to the core and enhancing the collision rate, which favours runaway growth and the formation of a massive central object.
	
	Comparing our stars-only results with those of \citet{2023MNRAS.522.4224V}, we find differences in the most massive object (MMO; our BH candidate) mass ranging from 0\% to 36\% (or 0\% to 6\% of the initial stellar cluster mass), with the largest discrepancies occurring in the most collisional models (A$_0$, B$_0$, and C$_0$). For family models A, B, C, and H, which lie in the IR, the MMO masses for stars-only variants (i.e., A$_0$, B$_0$, C$_0$ and H$_0$) are $13\,300$, $5\,250$, $1\,920$, and $12\,100~M_{\odot}$, respectively, placing them in the IMBH mass range ($\sim10^2$--$10^5~M_{\odot}$). Although models L$_0$ and M$_0$ are in the SR, their MMO masses are $64\,000$ and $5\,510~M_{\odot}$, respectively. This is because models L$_0$ and M$_0$ are among the most massive clusters; despite being in the SR, their large total mass still allows the formation of an IMBH. In contrast, they show lower efficiencies, meaning that the dynamics are controlled by the star cluster. If we compare models with the same initial mass, we observe that larger $M/M_{\rm crit}$ corresponds to higher MMO mass and BH formation efficiency. For example, models C$_0$, I$_0$, and N$_0$, all with an initial mass of $10^4~M_{\odot}$, form MMOs with masses of $1\,920$, $980$, and $3~M_{\odot}$, and efficiencies of 43.84\%, 17.82\%, and 0.03\%, respectively. This suggests that more massive clusters (potentially in the UR) could form even more massive central objects, with $\varepsilon_{\rm BH} \sim 1$.
	
	The gas can affect the cluster dynamics through two distinct mechanisms. When treated as a static background potential, the gas increases the stellar velocities, causing a delay in core collapse. As shown by \citet{reinoso2020effects}, this delay also affects the runaway growth, although for sufficiently long times the BH formation is enhanced, because the cluster maintains its compactness. On the other hand, a realistic gas can exert DF on stars through gas-star interactions \citep{ostriker1999dynamical, reinoso2020effects,leigh2014stellar}. In our simulations, the stellar velocity dispersion is transonic ($\sigma \gtrsim c_{\rm s}$), as shown in Figs.~\ref{fig:cs_profile_gas_A} and~\ref{fig:cs_profile_gas_L}. In this regime, DF is expected to be particularly efficient \citep{ostriker1999dynamical}. While our simulations do not fully resolve the DF (see Appendix~\ref{app:dynamical_friction}), the dissipation of kinetic energy due to gas-star interactions is still present. Consistent with this, our simulations show that the efficiencies are enhanced, with increases of $12.56 \pm 16.27\%$, $27.01 \pm 20.79\%$, and $45.10 \pm 31.39\%$ relative to the stars-only simulations for gas fractions of 5\%, 10\%, and 20\%, respectively. For the MMO mass, the corresponding increases are $8.40 \pm 15.89\%$, $22.51 \pm 20.57\%$, and $44.14 \pm 21.32\%$. We also find shorter core collapse times. On average, the core collapse time is reduced to $93.77 \pm 18.36\%$, $84.48 \pm 12.32\%$ and $72.61 \pm 8.99\%$ relative to the stars-only variants for gas fractions of 5\%, 10\% and 20\%, respectively. These results show that IMBH candidates can form in less massive clusters if a gaseous component is present. Although we use a modified critical mass relation compared with \citet{2023MNRAS.522.4224V}, the trend of increasing BH efficiency with $M/M_{\rm crit}$ is preserved. The fitted function (Eq.~\eqref{eq:ebh_model}), which includes the gas dependence through $x_{\rm eff}$, captures this trend.
	\subsection{Implications for SMBH formation}
	We have shown that in our simplified models, the formation of a BH candidate in the IMBH range is possible, with gas playing an important role by enhancing the number of collisions and, consequently, the final mass of the MMO. However, an important question is how massive a BH can become in more realistic NSCs. The sizes and masses of NSCs have been studied extensively \citep{neumayer2020nuclear}, with masses ranging from $10^5$ to $10^8~M_{\odot}$. Using the scaling relations from \citet{georgiev2016masses}, a nuclear star cluster of mass $M \sim 10^7~M_{\odot}$ has a minimum effective radius of $\sim 4$ pc. We estimate the critical mass ratio $M/M_{\rm crit}$ for a cluster of equal-mass stars ($m=1~M_{\odot}$, $r=1~R_{\odot}$), finding $M/M_{\rm crit}=4.3 \times 10^{-5}$ for $t_{\rm H}=5$ Myr. Using Eq.~\eqref{eq:ebh_model}, the predicted BH efficiency is $\sim 0.014$. Since $M/M_{\rm relax} > 1$ and the escaped mass is negligible for these mass ratios (see Fig.~\ref{fig:eff_mesc_Mrate}), we approximate $M_{\rm MMO} \simeq \varepsilon_{\rm BH}M$, yielding $M_{\rm MMO} \simeq 1.46\times10^5~M_{\odot}$. Including a gas fraction of $20\%$, $M/M_{\rm crit}=4.9 \times 10^{-5}$ and the efficiency increases to $\sim 0.033$, giving $M_{\rm MMO} \simeq 3.33\times10^5~M_{\odot}$. While the collisional and relaxation timescales are global quantities, the core of a nuclear star cluster could reach densities high enough to collapse into a massive BH, even as the outer stars expand due to relaxation processes \citep{2021ApJ...908...57E}. Since observed NSCs have expanded due to relaxation, they must have been more compact at formation. A more compact cluster has a higher $M/M_{\rm crit}$, and therefore a higher BH formation efficiency. An intermediate efficiency allows the formation of a BH while retaining a substantial NSC, consistent with the observed coexistence of NSCs and SMBHs in intermediate-mass galaxies \citep{georgiev2016masses,kritos2023massive,2021ApJ...908...57E}. Simulations of the most compact clusters, with radius $R\leq 0.5$ pc and mass $10^7~M_{\odot}$, show that a million solar mass black hole can form within $100$ Myr \citep{kritos2023massive}. For such a system ($M=10^7~M_{\odot}$, $R=0.5$ pc) composed only of stars, we estimate an efficiency of $0.068$ for $t_{\rm H}=5$ Myr, giving an upper limit on the MMO mass of $6.8 \times 10^5~M_{\odot}$. For a gas fraction of $20\%$, the estimated upper limit is $7.0 \times 10^5~M_{\odot}$. Despite the simplifications, our estimates show that in more massive NSCs it is possible to form a BH seed of $\sim 10^5~M_{\odot}$ on timescales of a few Myr, with gas playing an important role.
	
	Several mechanisms have been proposed in the literature for the growth of IMBHs to masses in the SMBH range. \citet{pelupessy2007rapidly} simulated halos containing IMBHs formed from Population III remnants and found that in collapsing halos with $M_{\rm halo} \lesssim 10^{10}~M_{\odot}$, accretion at the Eddington rate is not achieved due to stellar feedback. Even in the most massive halos, the growth of SMBH seeds is driven primarily by BH--BH mergers. They propose that the BH mass scales with the host halo mass as $M_{\rm BH} = 5 \times 10^6~M_{\odot}(M_{\rm halo}/10^{11}~M_{\odot})^{0.78}$. In contrast, \citet{shi2024seeds} describe a scenario in which subclusters and IMBHs form within a gas cloud complex of mass $10^8~M_{\odot}$. The hierarchical merging of subclusters occurs on a timescale of $\lesssim 1$ Myr, and the BHs in the subclusters migrate to the centre within the same period. The formation of a disc enables large accretion rates ($\sim 1$--$10~M_{\odot} {\rm yr}^{-1}$), allowing an IMBH to grow to an SMBH of $\gtrsim 2 \times 10^6~M_{\odot}$ within $1$ Myr. \citet{chung2026dense} showed through simulations that an IMBH can form in dense stellar clusters, with masses ranging from $\sim 300$ to $\sim 5000~M_{\odot}$ via runaway collisions. The resulting BHs can grow to $\sim 6 \times 10^3~M_{\odot}$ on a timescale of $\sim 10$ Myr through Eddington-limited gas accretion and TDEs. If a steady gas supply is assumed, with continuous gas accretion at the Eddington rate and a constant TDE rate, the BHs can reach masses of $\sim 6.2 \times 10^4~M_{\odot}$ on a timescale of $100$ Myr, resulting in promising SMBH seeds. These mechanisms are consistent with the formation of IMBHs in dense NSCs (such as those produced in our simulations) and suggest that gas accretion and IMBH mergers provide a viable pathway for seeds to grow to the SMBH regime.
	\subsection{Caveats}
	\label{Disc:caveats}
	In this section, we summarize the simplifications made in our simulations, which should be taken into account for improvements in future work. As we described in Section~\ref{sec:gas_dynamics}, the gas is evolved using an SPH method, with self-gravity and hydrodynamics, but without radiative cooling or heating. Radiative cooling allows the gas to lose thermal energy, which can help it remain gravitationally bound for longer periods. As shown by \citet{silich2018gas}, in sufficiently massive and compact clusters, radiative cooling can prevent the development of a global star cluster wind, thus suppressing gas expulsion. Moreover, recent work by \citet{schleicher2026massive} suggests that in the most compact and massive young clusters detected by JWST ($M \gtrsim 6 \times 10^6 M_{\odot}$), supernova feedback may be insufficient to expel the gas, allowing it to be retained and potentially enhance both DF and Bondi accretion onto the central object. The findings of \citet{schleicher2026massive} suggest that the gas-driven enhancement of runaway collisions we observe may be even stronger in more realistic scenarios. In addition, gas accretion could contribute to enhancing BH formation, since it increases the stellar density \citep{bonnell2002accretion,reinoso2018collisions}. On the other hand, stellar winds and supernovae (SNe) contribute to the chemical enrichment of the remnant gas after star formation episodes. Although the gas is heated, it does not necessarily evaporate from the cluster. As shown by \citet{silich2016gas}, depending on the gas density, the remnant gas can be expelled from the cluster or remain, in which case it can contribute to the runaway collision process. To assess the realism of our simplifications, it is important to consider the timescales of the relevant processes. Runaway collisions begin near core collapse ($t_{\rm cc}$), and a massive object forms on approximately the collisional timescale $t_{\rm coll}$. Any process that evaporates the gas on a timescale $t \lesssim  t_{\rm cc}$ has an important impact on the runaway collision process, while for timescales $t \gtrsim t_{\rm coll}$, the runaway process is not significantly perturbed. In an intermediate regime, $t_{\rm cc} \leq t \leq t_{\rm coll}$, the influence may be relevant.
	
	Some studies have investigated the timescales for gas expulsion in embedded clusters. \citet{dinnbier2020fast} studied gas expulsion from embedded star clusters, where the stellar mass is one third of the total mass (i.e., a gas fraction of $q=2$). Their results show that for clusters with total mass $M_{\rm cluster} \lesssim 3 \times 10^3~M_{\odot}$, the gas expulsion timescale is $t_{\rm ge} \gtrsim 0.9 $ Myr primarily through ionizing radiation and stellar winds. This mass range corresponds to models D, F, J, and K, for which $t_{\rm coll}$ (stars-only) is $1.018$, $10.829$, $27.032$, and $61.53$ Myr, respectively. We note that model D, which is in the IR, would evolve with a gas component in a more realistic scenario, at least during the runaway collision process. For more massive clusters, only $10\%$ to $30\%$ of the total mass could become unbound, and the majority could be ionized on a timescale of $\sim 1$ Myr. In contrast, \citet{calura2015feedback} ran simulations of globular clusters with a total mass of $ \sim 10^7~M_{\odot}$. Their hydrodynamical simulations included stellar feedback (winds and SNe), radiative cooling and self-gravity. They found that after $3$ Myr, stellar feedback removes $40\%$ of the initial gas, and after $14$ Myr, $99\%$ is removed. Similarly, \citet{pelupessy2012evolution} showed that for relatively small embedded clusters, the gas could be expelled by different mechanisms. While high feedback coupling could remove gas on a timescale of $\sim 5$ Myr, low feedback coupling is not sufficient to expel the gas via stellar winds, and SNe become the dominant mechanism (their first SN events occur at $\sim 10$ Myr). More recently, \citet{chung2026dense} showed that in hydrodynamical simulations of dense star cluster formation that include cooling, heating, and stellar feedback, the gas can be retained in sufficiently compact and massive clusters, affecting the IMBH formation process via runaway collisions. Taken together, these studies suggest that gas can persist during the core collapse phase in many realistic scenarios, particularly in low-mass clusters or with low stellar feedback. However, the survival of the gas depends on the details of gravity and stellar feedback, which we have not modelled. Therefore, although it seems plausible that gas could persist during an important part of the runaway collisions and could effectively enhance the formation of an IMBH, more realistic simulations that include stellar feedback, SNe, and radiative cooling are necessary.
	
	Our simulations do not include stellar evolution, and evolve for a timescale of 5 Myr, where most stars remain in the main sequence, before SNe. However, the MMO could collapse into a BH in shorter timescales. \citet{ohkubo2006core} showed that very massive stars (VMSs) of initially $500$ and $1000~M_{\odot}$ reach core-collapse on a timescale of $\sim 2$ Myr, producing BHs of $\sim 230$ and $\sim 500~M_{\odot}$, respectively. Simulations by \citet{chung2026dense} produce VMSs from $ 3.4 \times 10^2$ up to $5.1 \times 10^3~M_{\odot}$, which collapse into an IMBH in timescales from $2.58$ to $3.22$ Myr. Their models with VMSs over $\sim3\times10^3~M_{\odot}$ collapse directly into an IMBH with no mass loss. Considering a timescale of $2.5$ Myr, the MMO in most of our models has reached $\sim 90$--$100\%$ of their final mass, with the exceptions of L ($\gtrsim76\%$), M ($\gtrsim24\%$) and N (no VMS formed). In particular, at $2.5$ Myr, family models A, B, C, H, I, L and M form VMSs with masses $\gtrsim10^3~M_{\odot}$, which are above the pair-instability threshold and are therefore expected to collapse directly into an IMBH, retaining most of their mass \citep{heger2002nucleosynthetic}.
	
	There are numerical limitations that should be mentioned. Simulations of denser models in the UR would require a code more efficient than \textsc{PH4} in \textsc{AMUSE}, and capable of detecting stellar collisions. Also, our N-body code (\textsc{PH4}) does not include regularization for close encounters or a dedicated few-body treatment. Instead, close encounters are resolved by reducing the global timestep, which increases computational cost and may introduce errors in the orbits of hard binaries. Since binaries are a source of heat for the cluster, numerical errors in their formation or dissolution, as well as errors in collisions could bias the core collapse time and the number of collisions. Furthermore, as shown in Appendix~\ref{app:dynamical_friction}, DF is not fully resolved in our simulations, and the number of SPH particles within the gravitational radius is negligible. We therefore caution that gas DF is not fully captured. To achieve realistic DF, it would be necessary to increase the resolution of the SPH particles or to include a sub-grid model for unresolved gas DF, which could enhance the formation of an SMBH. As mentioned in \citet{pelupessy2012evolution} and \citet{whitehead2013simulating}, small differences in initial conditions are amplified due to the chaotic nature of N-body systems. Therefore, future work should perform multiple runs with different random realizations for each set of initial conditions. Although the general results will remain, this would allow us to obtain statistics for the trends shown in Section~\ref{sec:results}. Another simplification concerns the initial conditions. Stars are initialized using a Plummer distribution, with equal-mass stars in virial equilibrium. The gas is initialized using a Plummer distribution with the same spatial distribution as the stars. Although the Plummer distribution is consistent with observations \citep{plummer1911problem} and simulations \citep{chung2026dense}, it is necessary to include an IMF for the stellar mass distribution. On the other hand, cosmological simulations could provide more realistic initial conditions, including stellar and gas distributions and the remnant gas fraction.
	\section{Conclusions}
	\label{sec:conclusions}
	In this work, we have studied the role of gas in the collisional runaway scenario for SMBH seed formation in idealised dense stellar clusters. Our models are less massive and significantly more compact than typical observed NSCs, which allows us to test different dynamical regimes (SR and IR). By performing a series of simulations with coupled N-body and SPH codes via the \textsc{BRIDGE} scheme, we have shown that the presence of a gaseous component, even at small fractions (5\%–20\% of the stellar mass), significantly affects the formation of a massive central object.
	
	Our main findings are:
	\begin{enumerate}
		\item The gas shortens core collapse, particularly for gas fractions of 10\% and 20\%, with the strongest effect at 20\% (reduction of $27.39\% \pm 8.99\%$ relative to gas-free simulations). The effect is less robust at 5\% gas fraction ($6.23\% \pm 18.36\%$), suggesting that a minimum gas fraction is required for a significant effect.
		\item The presence of gas enhances the formation of a massive object. In our simulations, the MMO mass increases by $8.40\% \pm 15.89\%$, $22.51\% \pm 20.57\%$, and $44.14\% \pm 21.32\%$ for gas fractions of 5\%, 10\%, and 20\%, respectively, and the BH formation efficiency shows a similar trend. The enhancement is most pronounced in the models with the highest collision rates within the IR (A, B, C, H), where the density is highest and the gas can most effectively contribute to the increase of the MMO mass.
		\item The fitted model for BH formation efficiency shows that the effect of gas is most important in the intermediate range $10^{-4} \leq M/M_{\rm crit} \leq 10^5$ where the efficiency increases with gas fraction. Since gas dynamical friction is not fully resolved in our simulations, this enhancement may be a conservative estimate, and higher resolution simulations could reveal an even stronger effect.
	\end{enumerate}
	
	Our findings address the question of how gas affects the global instability that leads to runaway collisions and, eventually, to the formation of promising SMBH seeds. While our simulations do not include stellar evolution, the VMSs formed in our models would likely collapse into IMBHs on timescales of a few Myr (Section~\ref{Disc:caveats}). These IMBHs could serve as seeds and subsequently grow to SMBH masses via gas accretion and mergers, as suggested by recent studies \citep{inayoshi2022rapid,kritos2023massive,chung2026dense}. This extrapolation to more massive systems and longer timescales remains to be tested with simulations that include stellar evolution and feedback.
	
	We conclude that gas is a key ingredient in the collisional runaway scenario for SMBH seed formation, and its inclusion in future simulations is essential for a complete understanding of the formation of massive black holes in nuclear stellar clusters.
	\begin{acknowledgements}
		We gratefully acknowledge the Federal Ministry of Education and Research and the state governments of Chile for supporting this project as part of the funding of the Kultrun cluster hosted at the Departamento de Astronomía, Universidad de Concepción. B. Cuevas, A. Escala, M. C. Vergara and D. R. G. Schleicher acknowledge funding through CATA (ANID BASAL project FB21003). B. Reinoso acknowledges funding through the European Research Council via ERC Consolidator grant KETJU (no. 818930). Also D. R. G. Schleicher acknowledges the Alexander von Humboldt - Foundation, Bonn, Germany.
	\end{acknowledgements}
	%
	\bibliographystyle{bibtex/aa}
	\bibliography{references}
	
		
		
		
		
		
		
		
		
		

\begin{appendix}
	\section{Critical mass}
	\label{app:Mcrit}
	In this section, we present a solution of the implicit equations for the critical mass presented in \citet{2023MNRAS.522.4224V}, where a virialised system is assumed. We also consider an external potential (i.e., the gas component). We begin with the derivation of the collisional timescale. For a system with $N$ identical stars of mass $m$ and radius $r_{\star}$, velocity dispersion $\sigma$, total mass $M$ and characteristic radius $R$, the collision rate is given by \citep{2008gady.book.....B}:
	\begin{equation}
		\frac{1}{t_{\mathrm{coll}}} = 16\sqrt{\pi} n \sigma r_{\star}^2(1 + \Theta)~,
	\end{equation}
	where $\Theta$ is the Safronov number:
	\begin{equation}
		\Theta \equiv \frac{Gm}{2\sigma^2 r_{\star}} = 9.54 \frac{m}{M_{\odot}} \frac{R_{\odot}}{r_{\star}} \left(\frac{100~\mathrm{km}~\mathrm{s}^{-1}}{\sigma}\right)^2~,
	\end{equation}
	and $n=3M/(4 \pi R^3 m)$ is the number density. Defining the cross-section $\Sigma_0 = 16 \sqrt{\pi} r_{\star}^2 (1 + \Theta)$, the collision time can be rewritten as:
	\begin{equation}
		t_{\mathrm{coll}} = \frac{1}{n \sigma \Sigma_0}.
			\label{eq:app_tcoll}
	\end{equation}
	We consider a stellar system in virial equilibrium with characteristic radius $R$. In the absence of an external potential, the velocity dispersion satisfies $\sigma = \sqrt{GM/R}$. When the system is embedded in a gravitationally bound external potential of mass $M_{\mathrm{ext}}$ with the same distribution (same characteristic radius $R$), the modified dispersion velocity yields \citep{reinoso2020effects}:
	\begin{equation}
		\sigma_{\mathrm{ext}} = \sqrt{\frac{GM}{R}}(1+q)~, \label{eq:app_sigma_ext}
	\end{equation}
	where $q = M_{\mathrm{ext}}/M$ and $M + M_{\mathrm{ext}} = M(1+q)$ is the total mass. First, we replace the dispersion velocity (Eq.~\eqref{eq:app_sigma_ext}) into the cross-section $\Sigma_0$:
	\begin{align}
		\Sigma_0 =& 16 \sqrt{\pi} r_{\star}^2 (1 + \Theta) \nonumber \\
		=& 16 \sqrt{\pi} r_{\star}^2 \left(1 + 9.54 \frac{m}{M_{\odot}} \frac{R_{\odot}}{r_{\star}} \left(\frac{v_{100}}{\sigma_{\mathrm{ext}}}\right)^2 \right) \nonumber \\
		=& 16 \sqrt{\pi} r_{\star}^2 + \frac{9.54 \times 16 \sqrt{\pi} r_{\star} m R_{\odot} v_{100}^2}{M_{\odot}} \frac{R}{GM(1+q)^2}~, 
		\label{eq:app_Sigma_0_ext}
	\end{align}
	where $v_{100} = 100~\text{km}~\text{s}^{-1}$. From the relation $t_{\mathrm{coll}} = t_{\rm H}$, we place all mass-dependent terms on left side:
	\begin{align}
		t_{\rm H} =& t_{\mathrm{coll}} = \frac{1}{n \sigma_{\mathrm{ext}} \Sigma_0} \nonumber \\
		=& \frac{4 \pi R^3 m}{3M} \frac{\sqrt{R}}{\sqrt{GM} (1+q) \Sigma_0} \nonumber \\
		\Sigma_0 M^{3/2} =& \frac{4 \pi R^3 m}{3 t_{\rm H}} \frac{\sqrt{R}}{\sqrt{G} (1+q)}~.
		\label{eq:app_mass_eq_1}
	\end{align}
	Now, we replace Eq.~\eqref{eq:app_Sigma_0_ext} into Eq.~\eqref{eq:app_mass_eq_1}:
	\begin{align}
		16 \sqrt{\pi} r_{\star}^2 M^{3/2} +& \frac{9.54 \times 16 \sqrt{\pi} r_{\star} m R_{\odot} v_{100}^2R}{M_{\odot} G (1+q)^2} M^{1/2} = \frac{4 \pi m R^{7/2}}{3 G^{1/2}(1+q) t_{\rm H}} \nonumber \\
		M^{3/2} +& \frac{9.54 m R_{\odot} v_{100}^2R}{r_{\star} G M_{\odot} (1+q)^2} M^{1/2} = \frac{\sqrt{\pi} m R^{7/2}}{12 G^{1/2} r_{\star}^2 (1+q) t_{\rm H}}~.
	\end{align}
	In order to simplify, we define the variables:
	\begin{align}
		\alpha R =& A =  \frac{9.54 m R_{\odot} v_{100}^2}{r_{\star} G M_{\odot} (1+q)^2} R\\
		\beta R^{7/2} =& C = \frac{\sqrt{\pi} m}{12 G^{1/2} r_{\star}^2 (1+q) t_{\rm H}} R^{7/2}~.
	\end{align}
	Then, the equation for the mass is
	\begin{equation}
		M^{3/2} + A M^{1/2} - C = 0~.
		\label{eq:app_Ms_equation}
	\end{equation}
	\subsection{Solving for $M$}
	The critical mass can be obtained solving Eq.~\eqref{eq:app_Ms_equation}. To simplify, we use the change of variable $u^2 = M$ to have a cubic polynomial. Therefore, Eq.~\eqref{eq:app_Ms_equation} can be rewritten as:
	\begin{equation}
		u^3 + A u - C = 0~.
		\label{eq:app_Ms_to_u_eq}
	\end{equation}
	This equation could have three roots, but we are only interested in real solutions. Given $p = A/3$ and $r = -C$, the discriminant of Eq.~\eqref{eq:app_Ms_to_u_eq} is:
	\begin{equation}
		\Delta = 4p^3 + r^2 = \frac{4A^3}{27} + C^2~.
	\end{equation}
	Since $A, C >0$ it is straightforward that the discriminant is positive, then there is only one real solution \citep{zwillinger2002crc}:
	\begin{align}
		u_{\mathrm{real}} =& \sqrt[3]{s_1} + \sqrt[3]{s_2} \\
		s_1 =& \frac{-r + \sqrt{r^2 + 4p^3}}{2} \\
		s_2 =& \frac{-r - \sqrt{r^2 + 4p^3}}{2}~.
	\end{align}
	Replacing $p$ and $r$ by $ A/3$ and $-C$:
	\begin{align}
		u_{\mathrm{real}} =& \sqrt[3]{\frac{-r + \sqrt{r^2 + 4p^3}}{2}} + \sqrt[3]{\frac{-r - \sqrt{r^2 + 4p^3}}{2}} \nonumber \\
		=& \sqrt[3]{\frac{C}{2} + \sqrt{\frac{C^2}{4} + \frac{A^3}{27}}} + \sqrt[3]{\frac{C}{2} - \sqrt{\frac{C^2}{4} + \frac{A^3}{27}}}~.
	\end{align}
	Then, the critical mass is:
	\begin{align}
		M_{\mathrm{crit}} =& u_{\mathrm{real}}^2 \nonumber \\
		\begin{split}
		=& \left(\frac{C}{2} + \sqrt{\frac{C^2}{4} + \frac{A^3}{27}}\right)^{2/3} + \left(\frac{C}{2} - \sqrt{\frac{C^2}{4} + \frac{A^3}{27}}\right)^{2/3} +  \nonumber \\ & 2\sqrt[3]{\frac{C^2}{4} - \left(\frac{C^2}{4} + \frac{A^3}{27}\right)}
		\end{split} \\
		\begin{split}
		=& \left(\frac{C}{2} + \sqrt{\frac{C^2}{4} + \frac{A^3}{27}}\right)^{2/3} + \left(\frac{C}{2} - \sqrt{\frac{C^2}{4} + \frac{A^3}{27}}\right)^{2/3} -\\   
		&\frac{2A}{3}~,
		\end{split}
	\end{align}
	or in terms of $\alpha$ and $\beta$:
	\begin{equation}
		\begin{split}
		M_{\mathrm{crit}}=& \left(\frac{\beta R^{7/2}}{2} + \sqrt{\frac{\beta^2 R^7}{4} + \frac{\alpha^3 R^3}{27}}\right)^{2/3} + \\
		&\left(\frac{\beta R^{7/2}}{2} - \sqrt{\frac{\beta^2 R^7}{4} + \frac{\alpha^3 R^3}{27}}\right)^{2/3} -   
		\frac{2\alpha R}{3}~.
		\end{split} 
		\label{eq:app_Mcrit_sol}
	\end{equation}
	\subsection{Analytical vs numerical solution}
\begin{figure}[ht]
		\centering
		\includegraphics[width=0.5\textwidth]{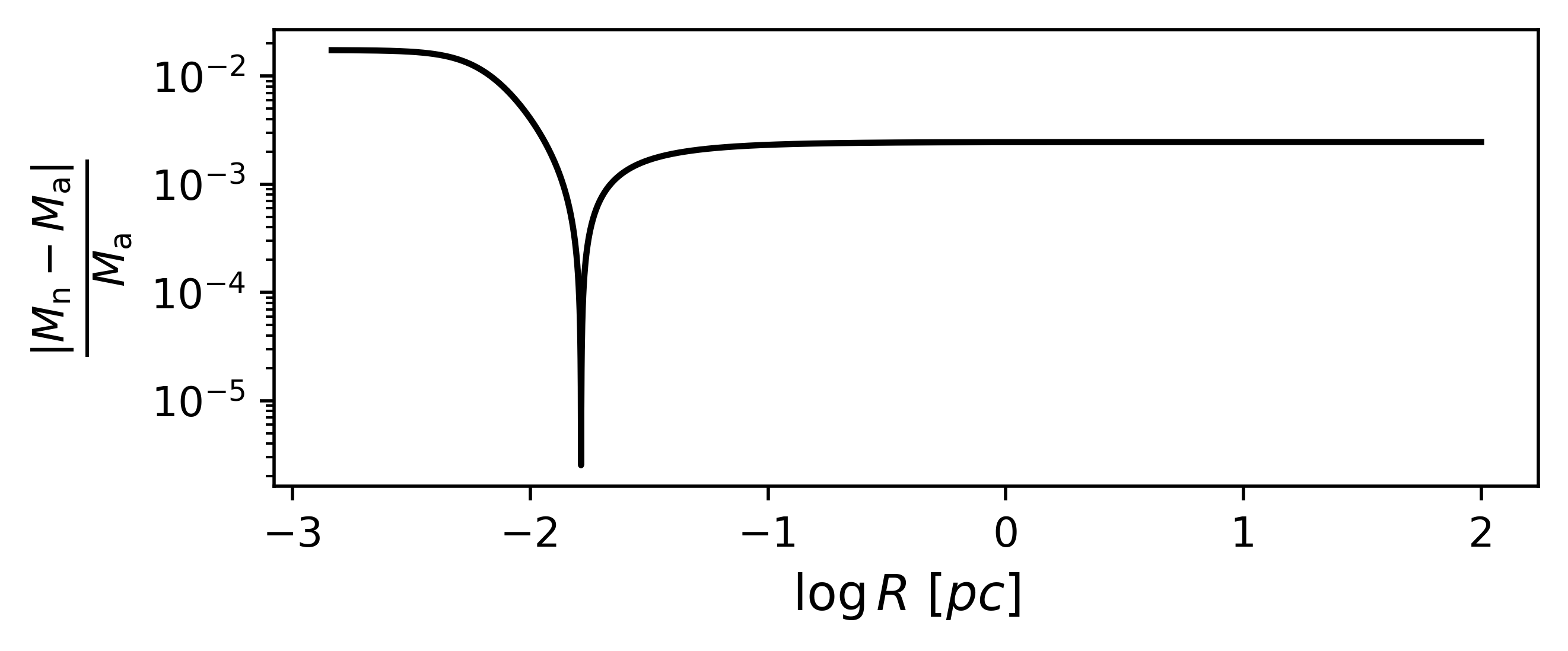}
		\caption{Relative error between the analytical solution ($M_{\mathrm{a}}$) and the numerical solution ($M_{\mathrm{n}}$) for the critical mass.}
		\label{fig:Mcrit_rel_err}
	\end{figure}

	To test the analytical solution for the critical mass, we compare it with a numerical solution obtained by solving the implicit condition iteratively. Fig.~\ref{fig:Mcrit_rel_err} shows the relative error between the numerical and analytical solutions as a function of the characteristic radius $R$.
	The relative difference between the numerical and analytical solutions is at most $\sim 1\%$ for small radii, decreasing to $\sim 0.1\%$ for larger radii. For all practical purposes, the analytical expression (Eq.~\eqref{eq:app_Mcrit_sol}) is accurate and numerically stable for $M_{\mathrm{crit}} \gtrsim 10~M_{\odot}$.
	\section{Relaxation mass}
	\label{app:Mrelax}
	In this section we derive a characteristic mass from the two-body relaxation timescale, in the presence of an external potential. Consider a system of $N$ equal-mass stars of mass $m$, with characteristic radius $R$ and velocity dispersion $\sigma$. A star crossing the system suffers a mean-square velocity deflection given by \citep{2008gady.book.....B}:
	\begin{equation}
		\Delta v^2 \simeq 8N \left(\frac{Gm}{Rv}\right)^2 \ln \left(\frac{b_{\mathrm{max}}}{b_{\mathrm{min}}}\right)~,
	\end{equation}
	where $b_{\mathrm{max}},~b_{\mathrm{min}}$ are the maximum and minimum impact parameters. Usually, $b_{\mathrm{max}} \simeq R$, and $b_{\mathrm{min}} = 2Gm/v^2$ is the impact parameter for a $90°$ deflection. The number of crossings needed for the star to change its velocity significantly is therefore
	\begin{equation}
		n_{\mathrm{relax}} =\frac{v^2}{\Delta v^2} = \left(\frac{Rv^2}{Gm}\right)^2 \frac{1}{8N \ln (b_{\mathrm{max}}/b_{\mathrm{min}})}~.
	\end{equation}
	As shown in Appendix~\ref{app:Mcrit}, the velocity dispersion in the presence of an external potential of mass $M_{\mathrm{ext}}=q M$ (with the same spatial distribution) becomes $\sigma_{\mathrm{ext}} = \sqrt{GNm/R}\,(1+q)$, where $M=Nm$ is the stellar mass. Setting $v=\sigma_{\mathrm{ext}}$ we obtain $Rv^2/(Gm) = RGNm(1+q)^2/(RGm) = N(1+q)^2$ and
	\begin{align}
		\ln \left(\frac{b_{\mathrm{max}}}{b_{\mathrm{min}}}\right) =& \ln \left(\frac{R}{(2Gm/v^2)}\right) = \ln \left(\frac{Rv^2}{2Gm}\right) \nonumber \\
		=& \ln \left(\frac{N(1+q)^2}{2}\right) = \ln(N(1+q)^2) - \ln(2) \nonumber \\
		\simeq& \ln(N(1+q)^2)~,
	\end{align}
	where we have dropped the constant $\ln(2)$ for simplicity. Substituting into the expression for $n_{\mathrm{relax}}$ yields
	\begin{equation}
		n_{\mathrm{relax}} = \frac{N(1+q)^4}{8 \ln(N(1+q)^2)} \simeq \frac{0.1 N(1+q)^4}{ \ln(N(1+q)^2)}~.
	\end{equation}
	Since $n_{\mathrm{relax}}$ is the number of crossing times required for relaxation, the relaxation time is $t_{\mathrm{relax}}= n_{\mathrm{relax}} t_{\mathrm{cross}}$, where $t_{\mathrm{cross}}=R/\sigma_{\mathrm{ext}}$. Thus,
	\begin{equation}
		t_{\mathrm{relax}} = n_{\mathrm{relax}} t_{\mathrm{cross}} = \frac{0.1 N(1+q)^4}{ \ln(N(1+q)^2)} t_{\mathrm{cross}}~.
	\end{equation}
	We now impose that the characteristic time $t_{\rm H}$ is shorter than the relaxation time ($t_{\rm H} \le t_{\mathrm{relax}}$), which is the condition for the system to retain its initial properties (two‑body relaxation is not yet effective). Similar to \citet{2023MNRAS.522.4224V}, we obtain a lower bound on the cluster radius:
	\begin{align}
		t_{\rm H} \leq& \frac{0.1 N(1+q)^4}{\ln(N(1+q)^2)} t_{\mathrm{cross}} = \frac{0.1 N(1+q)^4}{ \ln(N(1+q)^2)} \frac{R}{\sigma_{\mathrm{ext}}}  \nonumber \\
		=& \frac{0.1 M(1+q)^4}{ m\ln(M(1+q)^2/m)} \frac{R^{3/2}}{\sqrt{GM}(1+q)} \nonumber \\
		=& \frac{0.1 M^{1/2}(1+q)^3}{\ln(M(1+q)^2/m)} \frac{R^{3/2}}{G^{1/2}m} \nonumber \\
		R \geq& \left(\frac{t_{\rm H} m}{0.1 (1+q)^3}\ln\left(\frac{M(1+q)^2}{m}\right)\right)^{2/3} \left(\frac{G}{M}\right)^{1/3}~.
		\label{eq:app_R_relax}
	\end{align}
	\subsection{Solving for the relaxation mass}
	\label{ap:solve_Mrelax}
\begin{figure}[ht]
		\centering
		\includegraphics[width=0.4\textwidth]{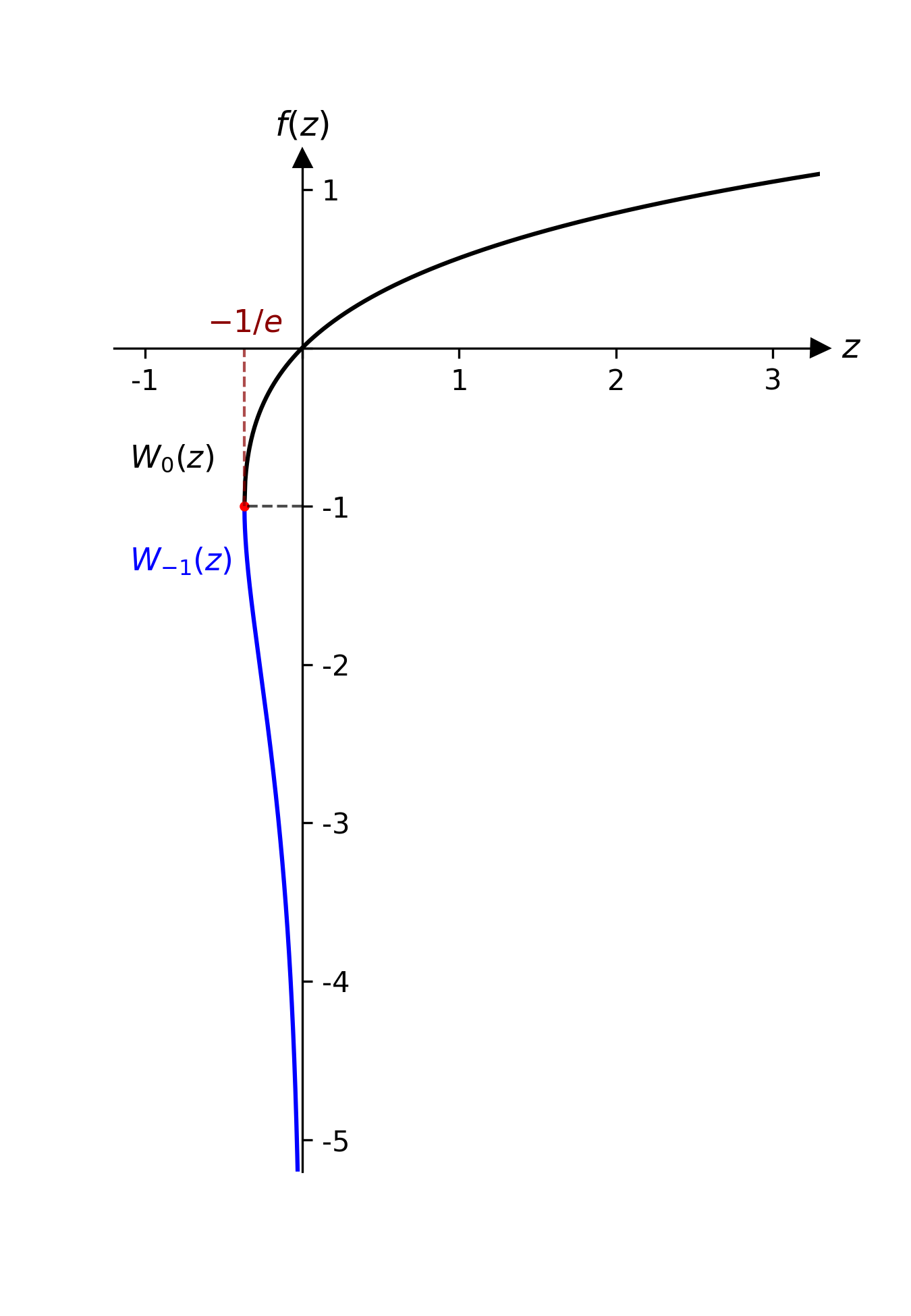}
		\caption{Lambert function. The black line shows the $W_{0}(z)$ branch, and the blue line shows the $W_{-1}(z)$ branch.}
		\label{fig:Lambert_function_branches}
	\end{figure}

	Inequality~\eqref{eq:app_R_relax} can be read as a relation $R(M)$. The relaxation mass is then defined as its inverse, $M_{\mathrm{relax}}(R)=R^{-1}(R(M))$. Setting the equality gives
	\begin{align}
		R =& \left(\frac{t_{\rm H} m\ln(M(1+q)^2/m)}{0.1 (1+q)^{3}}\right)^{2/3} \left(\frac{G}{M}\right)^{1/3} \nonumber\\
		R^3 =& \left(\frac{t_{\rm H} m \ln[M(1+q)^2/m]}{0.1 (1+q)^{3}}\right)^2 \left(\frac{G}{M}\right) \nonumber\\
		M^{1/2} R^{3/2} =& \frac{m t_{\rm H} \ln[M(1+q)^2/m]}{0.1 (1+q)^{3}} G^{1/2}~.
		\label{eq:app_B7}
	\end{align}
	Considering the change of variable $u = M(1+q)^2/m$, Eq.~\eqref{eq:app_B7} becomes
	\begin{align}
		\frac{(um)^{1/2}R^{3/2}}{1+q} =& \frac{m t_{\rm H} \ln(u)}{0.1 (1+q)^3} G^{1/2} \nonumber \\
		u^{1/2}=& \ln(u) \frac{t_{\rm H} (Gm)^{1/2}}{0.1 R^{3/2}(1+q)^2}~.
		\label{eq:app_u_12}
	\end{align}
	Defining $\lambda =  t_{\rm H} (Gm)^{1/2}/(0.1 R^{3/2}(1+q)^2)$, Eq.~\eqref{eq:app_u_12} yields:
	\begin{equation}
		u^{1/2} = \lambda \ln(u),
	\end{equation}
	and introducing the change of variable $\omega = \ln(u)$
	\begin{align}
		e^{\omega/2} =& \lambda \omega \nonumber \\
		-1/(2\lambda)=& (-\omega/2) e^{-\omega/2}~.
		\label{eq:app_w}
	\end{align}
	The inverse of the function $f(w)=w e^{w}$ is the Lambert function $W$ (i.e. $w = W(y)$ if $y = w e^{w}$) \citep{weisstein2002lambert}. Hence
	\begin{align}
		\frac{-\omega}{2} =& W\left(\frac{-1}{2\lambda}\right) \nonumber \\
		\omega =& -2 W\left(\frac{-1}{2\lambda}\right)~.
	\end{align}
	Returning to $u = e^{\omega}$ we obtain
	\begin{equation}
		u = \exp{\left[-2 W\left(\frac{-1}{2\lambda}\right)\right]}~,
	\end{equation}
	and finally the relaxation mass:
	\begin{equation}
		M_{\mathrm{relax}} = \frac{m}{(1+q)^2} \exp{\left[-2 W\left(\frac{-1}{2\lambda}\right)\right]}~.
		\label{eq:app_Mrelax}
	\end{equation}
	Eq.~\eqref{eq:app_Mrelax} involves the Lambert function $W$ evaluated at a negative argument $z = -1/(2\lambda)$ (since $\lambda>0$). There are two real-valued branches, $W_{0}(z)$ and $W_{-1}(z)$ \citep[Chapter~III.17]{higham2015princeton}. The principal branch, $W_{0}(z)$, is defined for $z \in [-1/e, \infty)$. The lower branch $W_{-1}(z)$ is defined on $[-1/e, 0)$. Since $-1/(2\lambda) < 0$ and assuming $-1/e \leq -1/(2\lambda)$, both branches can be taken. From Fig.~\ref{fig:Lambert_function_branches}, we see that for $z \in [-1/e, 0)$, the principal branch takes values in $[-1, 0]$, while $W_{-1}(z)$ takes values in $(-\infty, -1]$.
	
	Evaluating in Eq.~\eqref{eq:app_Mrelax}, the branch $W_{0}$ gives $m e^2 /(1+q)^2 \geq M_{\mathrm{relax}} > 0$, while $W_{-1}$ gives $m e^2 /(1+q)^2 \leq M_{\mathrm{relax}} < \infty$. As we expect for the relaxation mass to be $M_{\rm relax} \gg m$, the physically relevant solution is the lower branch. Then, the relaxation mass can be written as:
	\begin{equation}
		M_{\mathrm{relax}} = \frac{m}{(1+q)^2}\;
		\exp\!\left[-2\,W_{-1}\!\left(-\frac{1}{2\lambda}\right)\right]~.
		\label{eq:app_Mrelax_branch}
	\end{equation}
	A real solution exists only when \(z \ge -1/e\), i.e. $\lambda \ge e/2$. For typical star clusters ($N\gg 1$, $q\lesssim 1$, and $t_{\rm H}$ large), this condition is amply satisfied, so the expression is well‑defined and yields $M_{\mathrm{relax}}\gg m$.
	\section{Core collapse detection method}
	\label{app:cc_detection}
	In this section, we describe the implemented method for detecting the core collapse, using Lagrangian radii. The core collapse corresponds to the moment in which the inner Lagrangian radii reach their minimum \citep{di2022introducing,reinoso2020effects}, followed by an expansion of the outer regions. This process is expected to occur on a timescale of $\sim 12-19 ~ t_{\rm relax}$ \citep{quinlan1996time}, and can be observed from the evolution of Lagrangian radii. However, as the MMO reaches a mass comparable to that enclosed within a given Lagrangian radius, the evolution of that radius can be dominated by the motion of the MMO rather than by the contraction of the stellar core. To avoid this, we use the smallest Lagrangian radius that is not dominated by the most massive object. The algorithm proceeds as follows:
	\begin{enumerate}
		\item For a given Lagrangian radius $r_i(t)$, we select data within $t \leq 40 t_{\rm relax}$.
		\item We apply a Savitzky–Golay filter (implemented via the \texttt{savgol\_filter} function from \texttt{scipy.signal}) with a quadratic polynomial to smooth the data using a smoothing window of size $sw$.
		\item We define a sliding window of size $W = fN$, where $f$ is a parameter in the range $[0,1]$, and $N$ is the number of data points.
		\item For each window position, we perform two linear regressions: one on the first half of the data points in the window and another on the second half. We assign to the central time of the window the difference in slopes, computed as $\Delta m = m_{\rm right} - m_{\rm left}$.
		\item After running over all the selected data, the core collapse time is identified as the time at which $\Delta m$ reaches its maximum value. This corresponds to the point of maximum curvature change in the Lagrangian radius.
	\end{enumerate}
	The window size is varied between 10\% and 30\% of the selected data. To preserve sufficient resolution, the smoothing window was set empirically to $sw = W/5$. The method was validated by visual inspection of the Lagrangian radii for each simulation, confirming that the identified $t_{\rm cc}$ consistently matches the inflection point in the $r_j(t)$ curves.
	
	In our simulations, the ratio $t_{\rm cc}/t_{\rm relax}$ has an average value of $\sim 12$, consistent with the lower end of the range of $\sim 12-19 ~ t_{\rm relax}$. However, the most collisional models, such as A and B, have ratios of $\sim6.6$ and $\sim9.3$. This discrepancy can be explained by the fact that the estimate of \citet{quinlan1996time} is based on collisionless models. Collisions dissipate kinetic energy and can accelerate core collapse in dense systems, and contribute to mass segregation. In addition to this systematic effect, the stochastic nature of N-body simulations requires multiple realizations to obtain robust statistics.
	\section{Code setup and performance}
	\label{sec:code_setup}
	\subsection{Code parameters and initial conditions}
	\label{app:IC_and_parameters}
	Table~\ref{tab:code_params} summarizes the key parameters for the N-body code \textsc{PH4} and the SPH code \textsc{FI}, from the AMUSE framework. Parameters not listed were left at their default values.
	\begin{table*}[h]
		\centering
		\caption{Key runtime parameters for \textsc{PH4} and \textsc{FI}.}
		\label{tab:code_params}
		\begin{tabular}{l l l p{6cm}}
			\toprule
			Code & Parameter & Value & Description \\
			\midrule
			\textsc{PH4} & \texttt{epsilon\_squared} & $(0.05\,R_\odot)^2$ & Gravitational softening length squared. \\
			& \texttt{force\_sync} & \texttt{True} & Forces \texttt{evolve\_model} to synchronize all particles at the requested time. \\
			& \texttt{timestep\_parameter} & $0.01$ & Controls the accuracy of the Hermite integrator. \\
			\midrule
			\textsc{FI} & \texttt{eps\_is\_h\_flag} & \texttt{True} & Sets gravitational softening equal to the SPH smoothing length. \\
			& \texttt{timestep} & $1.0~t_{\rm nb}$ & Initial timestep (in N-body time units). \\
			& \texttt{radiation\_flag} & \texttt{False} & Disables radiative cooling and heating. \\
			& \texttt{use\_hydro\_flag} & \texttt{True} & Enables SPH hydrodynamics. \\
			& \texttt{integrate\_entropy\_flag} & \texttt{False} & Evolves specific internal energy (not entropy). \\
			& \texttt{isothermal\_flag} & \texttt{False} & Disables isothermal equation of state. \\
			& \texttt{star\_formation\_flag} & \texttt{False} & Disables star formation. \\
			\bottomrule
		\end{tabular}
	\end{table*}
	Table~\ref{tab:initial_conditions} show the initial conditions of each model.
	\begin{table*}[h]
		\centering
		\caption{Initial conditions for stellar clusters. $R_v$ and $M$ are the virial radius and total mass of the stellar cluster, $r$ and $m$ are the initial radius and mass of stars, $N$ is the number of stars, $q$ is the gas fraction (gas mass divided by stellar mass), and $\sigma$ is the velocity dispersion. For models with gas, $M_{\rm SPH}$ is the total gas mass, $N_{\rm SPH}$ is the number of SPH particles, and $m_{\rm SPH}$ is the mass per SPH particle, with $m_{\rm SPH} = m/100$. For $q=0$, the gas-related columns are left blank.}
		\label{tab:initial_conditions}
		\begin{tabular}{clrrrrlrrlr}
			\toprule
			ID & $R_v [{\rm pc}]$ & $M [M_{\odot}]$ & $N$ & $m [M_{\odot}]$ & $r [R_{\odot}]$ & $q$ & $M_{\rm SPH} [M_{\odot}]$ & $N_{\rm SPH}$ & $m_{\rm SPH} [M_{\odot}]$ & $\sigma[{\rm km}~{\rm s}^{-1}]$ \\
			\midrule
			A$_0$ & 0.005 & $5 \times 10^4$ & $10^3$ & 50 & 11.7 & 0 &  &  &  & 146.66\\
			A$_5$ & 0.005 & $5 \times 10^4$ & $10^3$ & 50 & 11.7 & 0.05 & $2.5\times 10^3$ & $5\times 10^3$ & 0.5 & 146.66\\
			A$_{10}$ & 0.005 & $5 \times 10^4$ & $10^3$ & 50 & 11.7 & 0.1 & $5\times 10^3$ & $10^4$ & 0.5 & 146.66\\
			A$_{20}$ & 0.005 & $5 \times 10^4$ & $10^3$ & 50 & 11.7 & 0.2 & $10^4$ & $2 \times 10^4$ & 0.5 & 146.66\\
			\midrule
			B$_0$ & 0.005 & $2.5 \times 10^4$ & $5 \times 10^2$ & 50 & 11.7 & 0 &  &  &  & 103.71\\
			B$_5$ & 0.005 & $2.5 \times 10^4$ & $5 \times 10^2$ & 50 & 11.7 & 0.05 & $1.25\times 10^3$ & $2.5\times 10^3$ & 0.5 & 103.71\\
			B$_{10}$ & 0.005 & $2.5 \times 10^4$ & $5 \times 10^2$ & 50 & 11.7 & 0.1 & $2.5\times 10^3$ & $5\times 10^3$ & 0.5 & 103.71\\
			B$_{20}$ & 0.005 & $2.5 \times 10^4$ & $5 \times 10^2$ & 50 & 11.7 & 0.2 & $5\times 10^3$ & $10^4$ & 0.5 & 103.71\\
			\midrule
			C$_0$ & 0.005 & $10^4$ & $10^3$ & 10 & 4.7 & 0 &  &  &  & 65.59\\
			C$_5$ & 0.005 & $10^4$ & $10^3$ & 10 & 4.7 & 0.05 & $5 \times 10^2$ & $5 \times 10^3$ & 0.1 & 65.59\\
			C$_{10}$ & 0.005 & $10^4$ & $10^3$ & 10 & 4.7 & 0.1 & $10^3$ & $10^4$ & 0.1 & 65.59\\
			C$_{20}$ & 0.005 & $10^4$ & $10^3$ & 10 & 4.7 & 0.2 & $2 \times 10^3$ & $2 \times 10^4$ & 0.1 & 65.59\\
			\midrule
			D$_0$ & 0.005 & $5 \times 10^3$ & $5 \times 10^2$ & 10 & 4.7 & 0 &  &  &  & 46.38\\
			D$_5$ & 0.005 & $5 \times 10^3$ & $5 \times 10^2$ & 10 & 4.7 & 0.05 & $2.5 \times 10^2$ & $2.5 \times 10^3$ & 0.1 & 46.38\\
			D$_{10}$ & 0.005 & $5 \times 10^3$ & $5 \times 10^2$ & 10 & 4.7 & 0.1 & $5 \times 10^2$ & $5 \times 10^3$ & 0.1 & 46.38\\
			D$_{20}$ & 0.005 & $5 \times 10^3$ & $5 \times 10^2$ & 10 & 4.7 & 0.2 & $10^3$ & $10^4$ & 0.1 & 46.38\\
			\midrule
			F$_0$ & 0.005 & $10^3$ & $10^3$ & 1 & 1.0 & 0 &  &  &  & 20.74\\
			F$_5$ & 0.005 & $10^3$ & $10^3$ & 1 & 1.0 & 0.05 & $5 \times 10^1$ & $5 \times 10^3$ & 0.01 & 20.74\\
			F$_{10}$ & 0.005 & $10^3$ & $10^3$ & 1 & 1.0 & 0.1 & $10^2$ & $10^4$ & 0.01 & 20.74\\
			F$_{20}$ & 0.005 & $10^3$ & $10^3$ & 1 & 1.0 & 0.2 & $2 \times 10^2$ & $2 \times 10^4$ & 0.01 & 20.74\\
			\midrule
			H$_0$ & 0.01 & $5 \times 10^4$ & $5 \times 10^3$ & 10 & 4.7 & 0 &  &  &  & 103.71\\
			H$_5$ & 0.01 & $5 \times 10^4$ & $5 \times 10^3$ & 10 & 4.7 & 0.05 & $2.5 \times 10^3$ & $2.5 \times 10^4$ & 0.1 & 103.71\\
			H$_{10}$ & 0.01 & $5 \times 10^4$ & $5 \times 10^3$ & 10 & 4.7 & 0.1 & $5 \times 10^3$ & $5 \times 10^4$ & 0.1 & 103.71\\
			H$_{20}$ & 0.01 & $5 \times 10^4$ & $5 \times 10^3$ & 10 & 4.7 & 0.2 & $10^4$ & $10^5$ & 0.1 & 103.71\\
			\midrule
			I$_0$ & 0.01 & $10^4$ & $10^3$ & 10 & 4.7 & 0 &  &  &  & 46.38\\
			I$_5$ & 0.01 & $10^4$ & $10^3$ & 10 & 4.7 & 0.05 & $5 \times 10^2$ & $5 \times 10^3$ & 0.1 & 46.38\\
			I$_{10}$ & 0.01 & $10^4$ & $10^3$ & 10 & 4.7 & 0.1 & $10^3$ & $10^4$ & 0.1 & 46.38\\
			I$_{20}$ & 0.01 & $10^4$ & $10^3$ & 10 & 4.7 & 0.2 & $2 \times 10^3$ & $2 \times 10^4$ & 0.1 & 46.38\\
			\midrule
			J$_0$ & 0.01 & $5 \times 10^3$ & $5 \times 10^3$ & 1 & 1.0 & 0 &  &  &  & 32.79\\
			J$_5$ & 0.01 & $5 \times 10^3$ & $5 \times 10^3$ & 1 & 1.0 & 0.05 & $2.5 \times 10^2$ & $2.5 \times 10^4$ & 0.01 & 32.79\\
			J$_{10}$ & 0.01 & $5 \times 10^3$ & $5 \times 10^3$ & 1 & 1.0 & 0.1 & $5 \times 10^2$ & $5 \times 10^4$ & 0.01 & 32.79\\
			J$_{20}$ & 0.01 & $5 \times 10^3$ & $5 \times 10^3$ & 1 & 1.0 & 0.2 & $10^3$ & $10^5$ & 0.01 & 32.79\\
			\midrule
			K$_0$ & 0.01 & $10^3$ & $10^3$ & 1 & 1.0 & 0 &  &  &  & 14.66\\
			K$_5$ & 0.01 & $10^3$ & $10^3$ & 1 & 1.0 & 0.05 & $5 \times 10^1$ & $5 \times 10^3$ & 0.01 & 14.66\\
			K$_{10}$ & 0.01 & $10^3$ & $10^3$ & 1 & 1.0 & 0.1 & $10^2$ & $10^4$ & 0.01 & 14.66\\
			K$_{20}$ & 0.01 & $10^3$ & $10^3$ & 1 & 1.0 & 0.2 & $2 \times 10^2$ & $2 \times 10^4$ & 0.01 & 14.66\\
			\midrule
			L$_0$ & 0.1 & $5 \times 10^5$ & $10^4$ & 50 & 11.7 & 0 &  &  &  & 103.71\\
			L$_5$ & 0.1 & $5 \times 10^5$ & $10^4$ & 50 & 11.7 & 0.05 & $2.5 \times 10^4$ & $5 \times 10^4$ & 0.5 & 103.71\\
			L$_{10}$ & 0.1 & $5 \times 10^5$ & $10^4$ & 50 & 11.7 & 0.1 & $5 \times 10^4$ & $10^5$ & 0.5 & 103.71\\
			L$_{20}$ & 0.1 & $5 \times 10^5$ & $10^4$ & 50 & 11.7 & 0.2 & $10^5$ & $2 \times 10^5$ & 0.5 & 103.71\\
			\midrule
			M$_0$ & 0.1 & $10^5$ & $10^4$ & 10 & 4.7 & 0 &  &  &  & 46.38\\
			M$_5$ & 0.1 & $10^5$ & $10^4$ & 10 & 4.7 & 0.05 & $5 \times 10^3$ & $5 \times 10^4$ & 0.1 & 46.38\\
			M$_{10}$ & 0.1 & $10^5$ & $10^4$ & 10 & 4.7 & 0.1 & $10^4$ & $10^5$ & 0.1 & 46.38\\
			M$_{20}$ & 0.1 & $10^5$ & $10^4$ & 10 & 4.7 & 0.2 & $2 \times 10^4$ & $2 \times 10^5$ & 0.1 & 46.38\\
			\midrule
			N$_0$ & 0.1 & $10^4$ & $10^4$ & 1 & 1.0 & 0 &  &  &  & 14.67\\
			N$_5$ & 0.1 & $10^4$ & $10^4$ & 1 & 1.0 & 0.05 & $5 \times 10^2$ & $5 \times 10^4$ & 0.01 & 14.67\\
			N$_{10}$ & 0.1 & $10^4$ & $10^4$ & 1 & 1.0 & 0.1 & $10^3$ & $10^5$ & 0.01 & 14.67\\
			N$_{20}$ & 0.1 & $10^4$ & $10^4$ & 1 & 1.0 & 0.2 & $2 \times 10^3$ & $2 \times 10^5$ & 0.01 & 14.67\\
			\bottomrule
		\end{tabular}
	\end{table*}
	\subsection{Adaptive BRIDGE timestep}
	\label{app:Adaptive_BRIDGE_Timestep}
	The \textsc{BRIDGE} and \textsc{FI} codes use the same timestep, which is adjusted adaptively based on the local dynamical conditions of the gas and stars. Here, $t_{\rm nb}$ denotes the N-body time unit, and $t_{\rm relax}$ is the two-body relaxation time. The algorithm proceeds as follows:
	\begin{enumerate}
		\item The minimum free-fall time of the gas is computed as:
		\begin{equation}
			t_{\rm ff} = \sqrt{\frac{3\pi}{32G\rho_{\rm max}}},
		\end{equation}
		where $\rho_{\rm max}$ is the maximum density among SPH particles.
		\item A Courant-type timestep is computed from the accelerations between gas and star particles:
		\begin{equation}
			\Delta t_{\rm coup} = \min\left(
			f_{\rm coup}\sqrt{\frac{2h}{\mathbf{a}_{\rm g,s}}},
			f_{\rm coup}\sqrt{\frac{\epsilon_{\rm coup}}{\mathbf{a}_{\rm s,g}}}
			\right)~,
		\end{equation}
		where $h$ is the SPH smoothing length, $\epsilon_{\rm coup}$ is the coupling softening length, and $\mathbf{a}_{\rm g,s}$ and $\mathbf{a}_{\rm s,g}$ are the accelerations of gas and star particles due to the other component.
		\item The reference timescale is then defined as:
		\begin{equation}
			\tau = \min(t_{\rm ff}, \Delta t_{\rm coup})~.
		\end{equation}
		\item The BRIDGE timestep is adjusted so that it satisfies:
		\begin{equation}
			\frac{\tau}{2} \leq \Delta t_{\rm BRIDGE} \leq \tau~.
		\end{equation}
		If $\Delta t_{\rm BRIDGE} < \tau/2$, the timestep is multiplied by 2.
	\end{enumerate}
	To maintain synchronization with the N-body integrator, the BRIDGE timestep is rounded to the nearest power of two of the N-body time ($t_{\rm nb}$):
	\begin{equation}
		\Delta t_{\rm BRIDGE} = 2^{k} t_{\rm nb}~,
	\end{equation}
	with $k$ chosen such that the timestep lies within the range $[\tau/2, \tau]$. During the first 25 relaxation times ($t_{\rm relax}$), the maximum timestep is reduced to $2^{-4}~t_{\rm nb}$ to improve temporal resolution during the core collapse phase. After this period, the timestep may grow up to the initial value of $1.0~t_{\rm nb}$. For all simulations, $f_{\rm coup}=0.5$ and $\epsilon_{\rm coup} = 5~{\rm AU}$.
	\subsection{Dynamical friction}
	\label{app:dynamical_friction}
	To verify that dynamical friction is well resolved, we check two conditions: (i) the gravitational radius should be larger than the coupling softening length ($\epsilon_{\rm coup}$ parameter); and (ii) the number of SPH particles within the gravitational radius for a typical star should be $\gtrsim 100$. The gravitational radius is given by
	\begin{equation}
		R_{\rm g} = GM/(c_{\rm s}^2 + v^2)~, 
	\end{equation}
	where $M$ is the mass of an individual star, $c_{\rm s}$ is the gas sound speed, and $v$ is the velocity of the star. Among all models, the gravitational radius ranges between $\sim 0.1~{\rm AU}$ and $\sim 10~{\rm AU}$. The first condition is only marginally satisfied for the most massive stars, and is not satisfied for lower-mass stars. Furthermore, the number of SPH particles within $R_{\rm g}$ is effectively zero for all models. We therefore conclude that dynamical friction from the gas is not fully resolved in our simulations. However, this implies that the gas-driven enhancement of core collapse and BH formation observed in our simulations would be even stronger if dynamical friction were fully resolved. Future work with higher resolution would be required to quantify this effect accurately.
	\subsection{Energy conservation and virialisation}
\begin{figure}[h]
		\centering
		\includegraphics[width=0.5\textwidth]{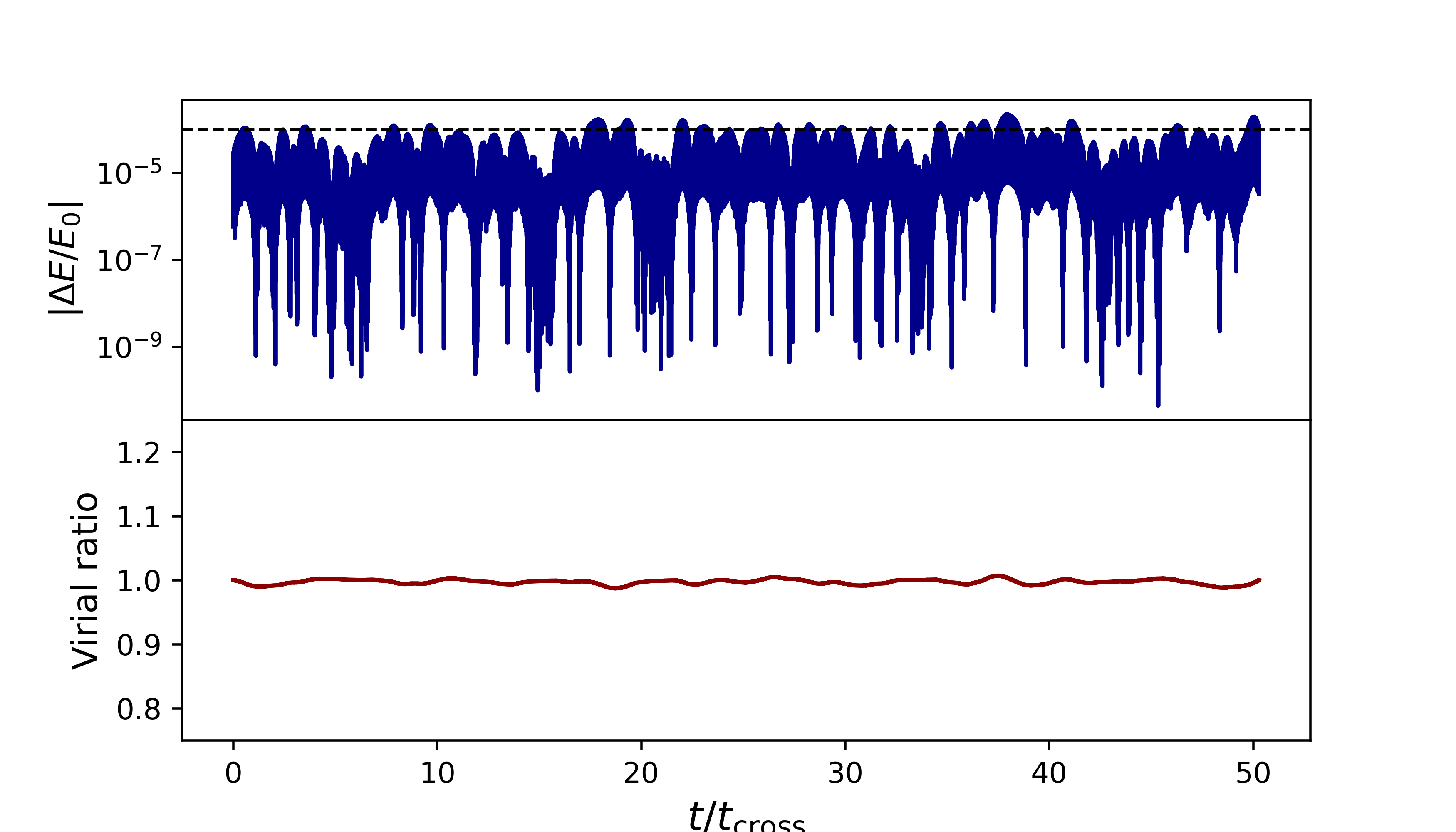}
		\caption{Numerical accuracy test for a system of 5000 stars of $1~M_{\odot}$ embedded in a gas cloud with the same mass and 5000 SPH particles. Top panel: relative error in total energy per timestep as a function of crossing time. Bottom panel: virial ratio. The system remains virialised during the first 50 crossing times, with an energy error $\lesssim 10^{-4}$.}
		\label{fig:Energy_cons}
	\end{figure}

	The validation of our results relies on the accuracy of energy conservation. Fig.~\ref{fig:Energy_cons} shows, in the top panel, the error in total energy per timestep as a function of crossing time. The bottom panel shows the virial ratio, defined as the ratio of two times the total kinetic energy (including thermal energy of gas) to the absolute value of the gravitational potential energy, $2K/|U|$. The initial conditions for this test consist of a stellar system with virial radius $R_v=1~{\rm pc}$, composed of 5000 stars of $1~M_{\odot}$ each, embedded in a gas cloud with the same virial radius, a total gas mass of $5000~M_{\odot}$, and containing 5000 SPH particles. We observe that during the first 50 crossing times, the system remains virialised, with an energy error of $\lesssim 10^{-4}$.
	\section{Extra plots}
	\label{app:extra_plots}
	In this appendix, we provide supplementary figures that support the discussion in Sections~\ref{sec:3_num_methods} and~\ref{sec:Discussion}.
	\subsection{Lagrangian radii of gas}
\begin{figure}[h]
		\centering
		\includegraphics[width=0.45\textwidth]{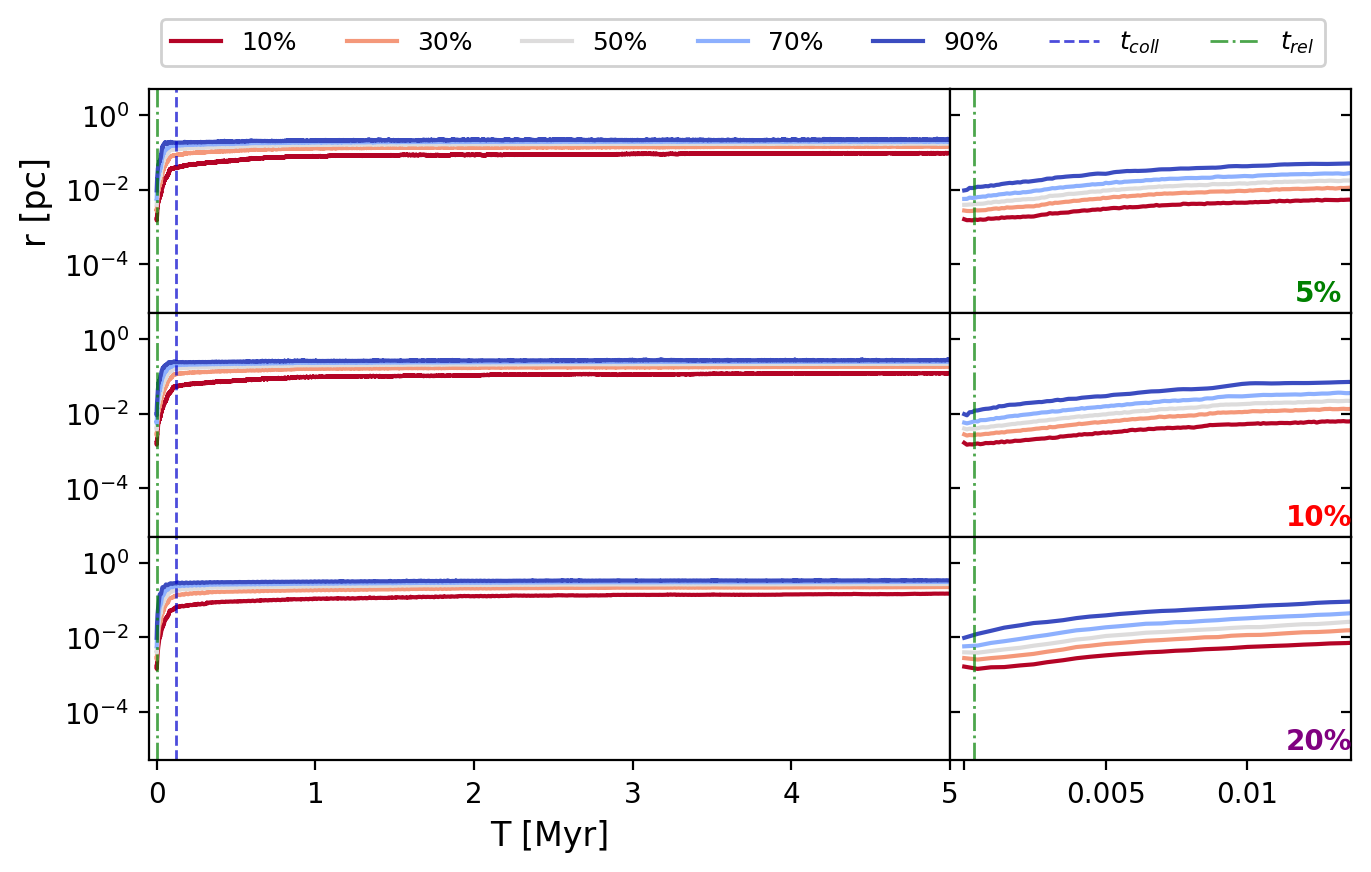}
		\caption{Lagrangian radii for gas of three variants of Model A. Top panel is for $\mathrm{A}_{5}$, middle is for $\mathrm{A}_{10}$ and bottom panel is for $\mathrm{A}_{20}$. Left column shows the full simulation time (5 Myr) while right column shows the evolution until 40 relaxation times.}
		\label{fig:Lrad_A_gas}
	\end{figure}

\begin{figure}[h]
		\centering
		\includegraphics[width=0.45\textwidth]{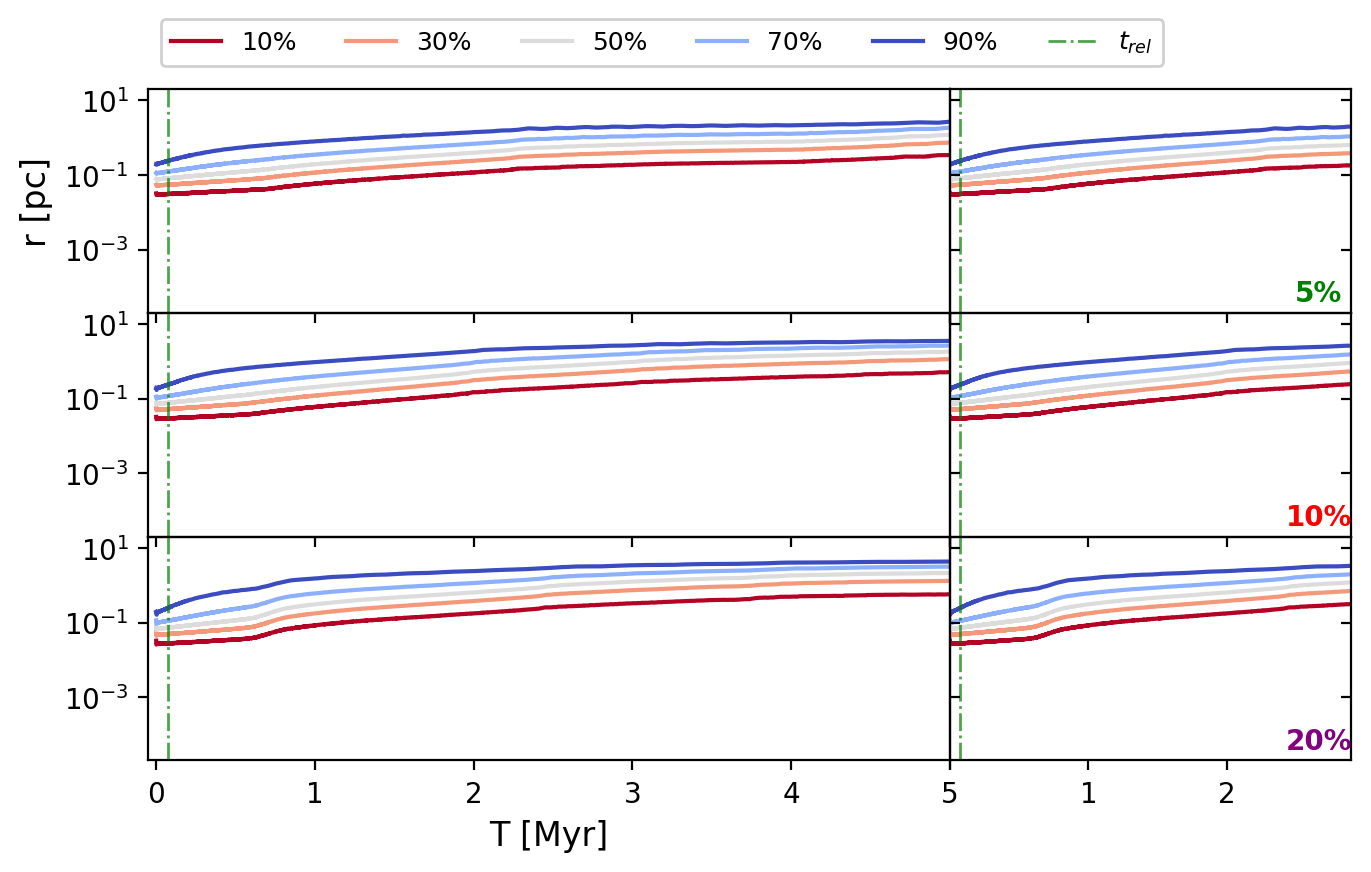}
		\caption{Lagrangian radii for gas of three variants of Model L. Top panel is for $\mathrm{L}_{5}$, middle is for $\mathrm{L}_{10}$ and bottom panel is for $\mathrm{L}_{20}$. Left column shows the full simulation time (5 Myr) while right column shows the evolution until 40 relaxation times.}
		\label{fig:Lrad_L_gas}
	\end{figure}

	The gas Lagrangian radii for the A and L families are shown in Figs.~\ref{fig:Lrad_A_gas} and~\ref{fig:Lrad_L_gas}, respectively.	
	\subsection{Sound speed}
\begin{figure}[h]
		\centering
		\includegraphics[width=0.5\textwidth]{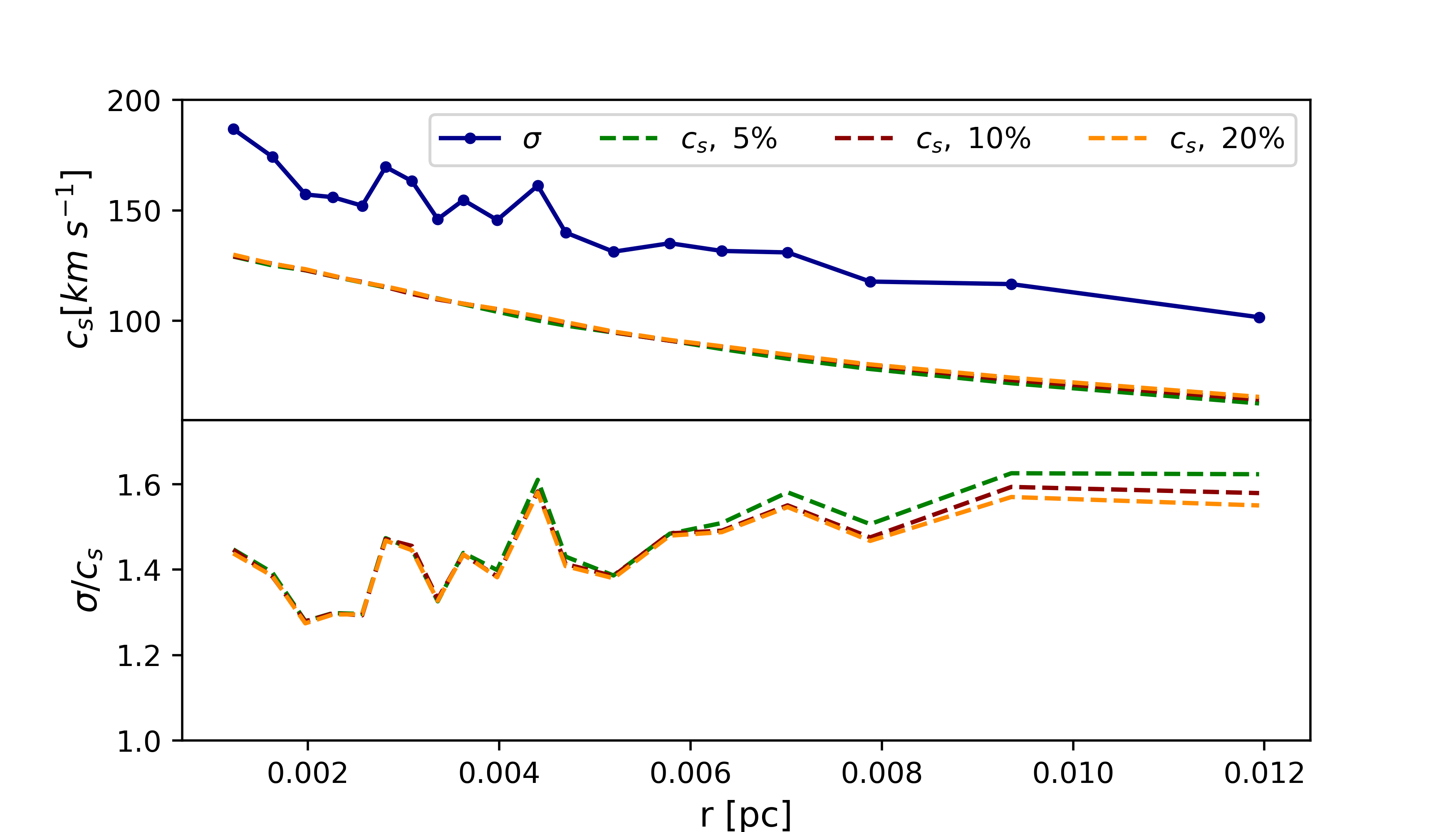}
		\caption{Sound speed of gas ($c_{\rm s}$) and velocity dispersion ($\sigma$) for family A models.}
		\label{fig:cs_profile_gas_A}
	\end{figure}

\begin{figure}[h]
		\centering
		\includegraphics[width=0.5\textwidth]{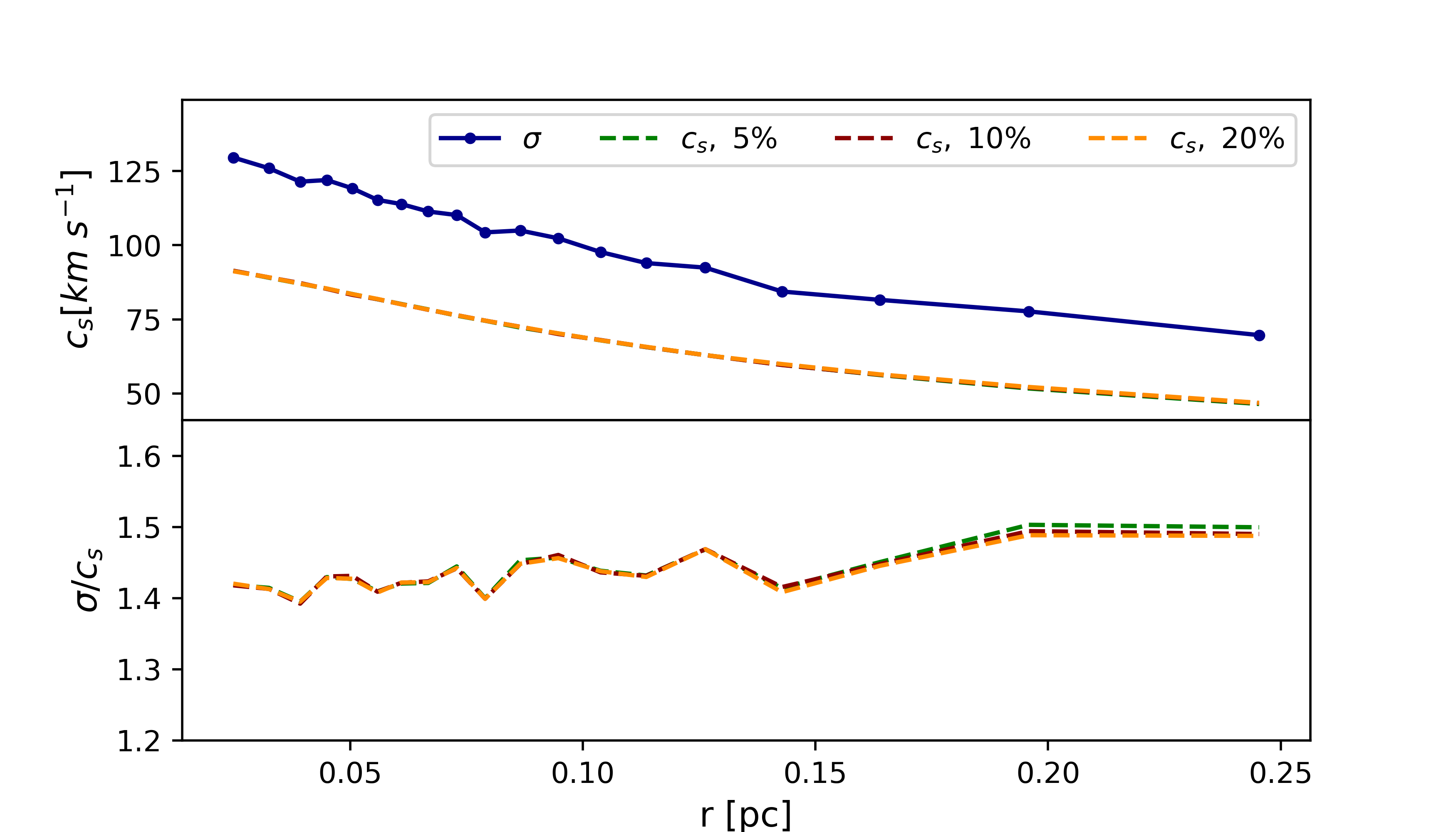}
		\caption{Sound speed of gas ($c_{\rm s}$) and velocity dispersion ($\sigma$) for family L models.}
		\label{fig:cs_profile_gas_L}
	\end{figure}

	The radial profiles of the gas sound speed and the stellar velocity dispersion are shown in Figs.~\ref{fig:cs_profile_gas_A} and \ref{fig:cs_profile_gas_L} for families A and L, respectively. The stellar velocity dispersion is transonic throughout the cluster for both family models, which is the regime where gas dynamical friction is expected to be most efficient.
	
	\section{Density of simulated clusters versus observations}
	\label{app:density_comparison}
\begin{figure}[h]
		\centering
		\includegraphics[width=0.45\textwidth]{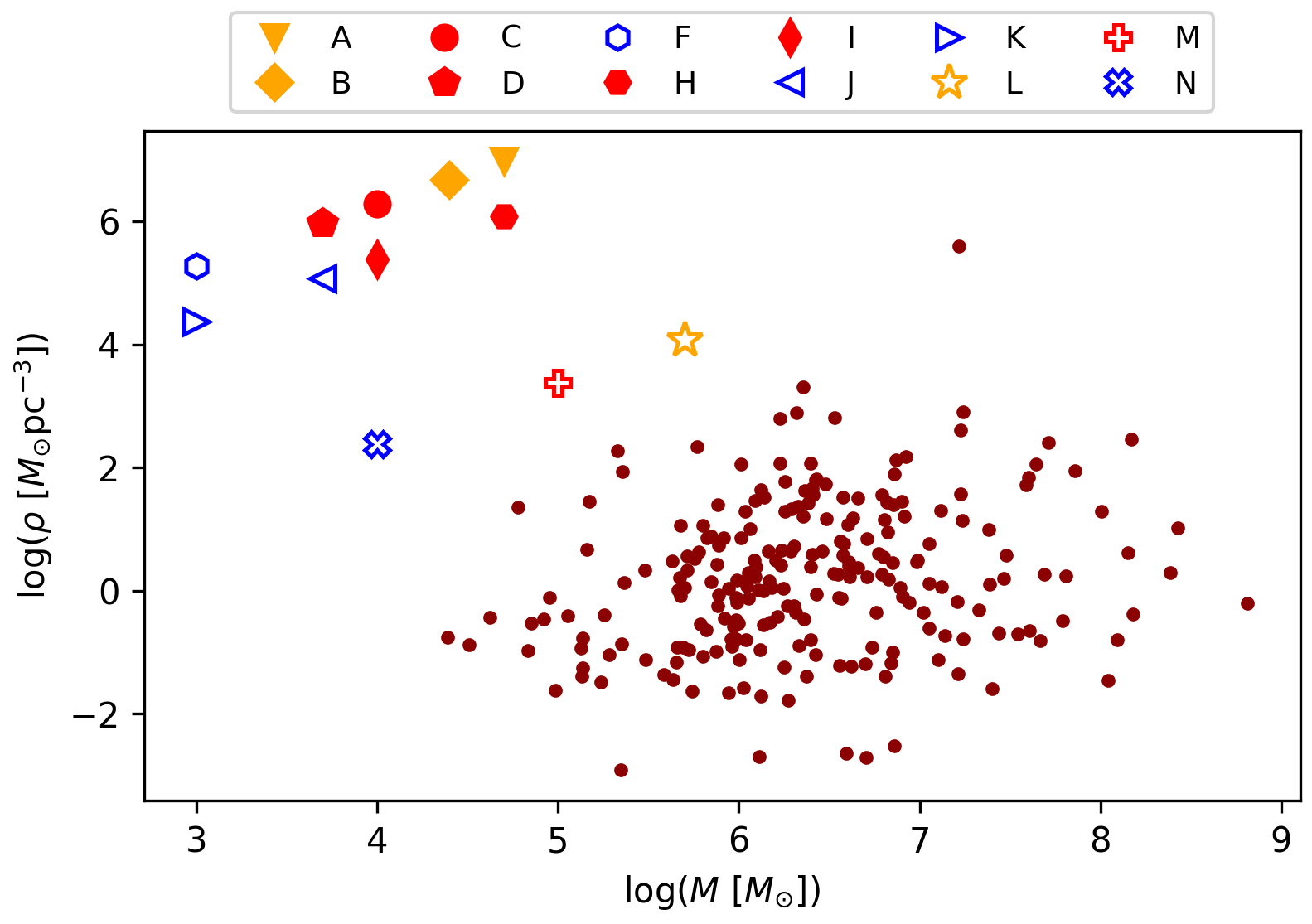}
		\caption{Density versus mass, on a logarithmic scale. The symbols correspond to the initial conditions of our stars-only models. Red circles are observational data of NSCs, taken from \citet{georgiev2016masses}.}
		\label{fig:dens_vs_mass}
	\end{figure}

	In this section, we compare the densities of our models with those of observed systems. We estimate the density of each system as:
	\begin{equation}
		\rho = \frac{3M}{4 \pi R^3}~,
	\end{equation}
	where $M$ is the stellar cluster mass and $R$ the characteristic radius. In Fig.~\ref{fig:dens_vs_mass} we show the estimated density for our stars-only models (symbols) and for observational data from \citet{georgiev2016masses} (red circles). For our models, we estimate the density using the virial radius and the gravitationally bound mass. For the observational data, we use the NSC mass and the effective radius.
	
\begin{figure}[h]
		\centering
		\includegraphics[width=0.5\textwidth]{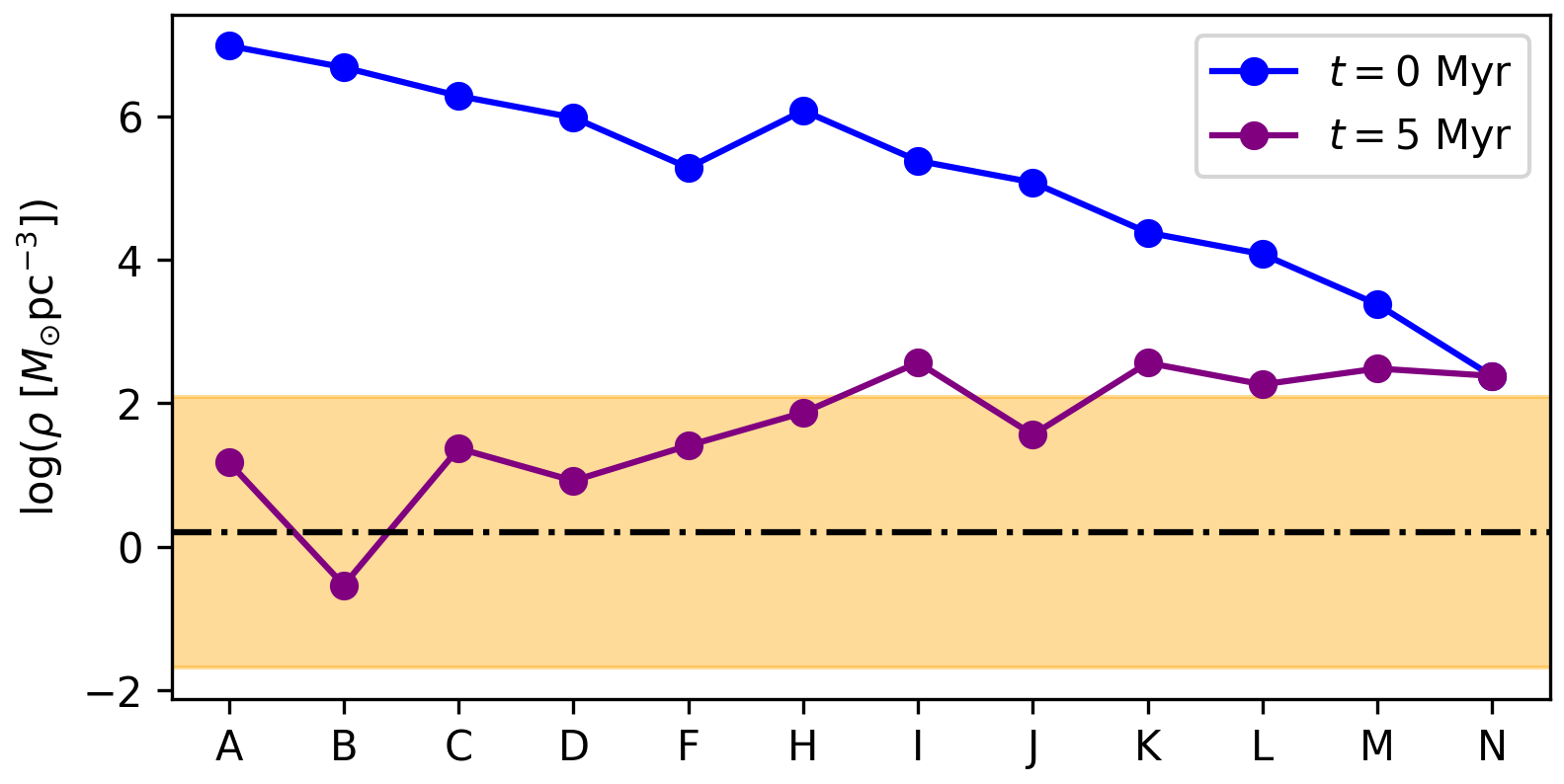}
		\caption{Density of our models on a logarithmic scale. Blue circles are for stars-only models at the beginning ($t=0~{\rm Myr}$), while purple circles are for the same models at the end of the simulation ($t=5~{\rm Myr}$). The dot-dashed black line represents the mean log-density of the observational data. The orange region indicates the $2\sigma$ range of the observational data.}
		\label{fig:dens_mean}
	\end{figure}

	In Fig.~\ref{fig:dens_mean} we show the density of our stars-only models at $t=0~{\rm Myr}$ (blue circles) and at $t=5~{\rm Myr}$ (purple circles). The dot-dashed black line represents the mean log-density of the observational data, with the orange region indicating the $2\sigma$ (with $\sigma$ the standard deviation) range. Our models are considerably denser than the observational data. However, as they evolve, their densities decrease.
	
\begin{figure}[h]
		\centering
		\includegraphics[width=0.5\textwidth]{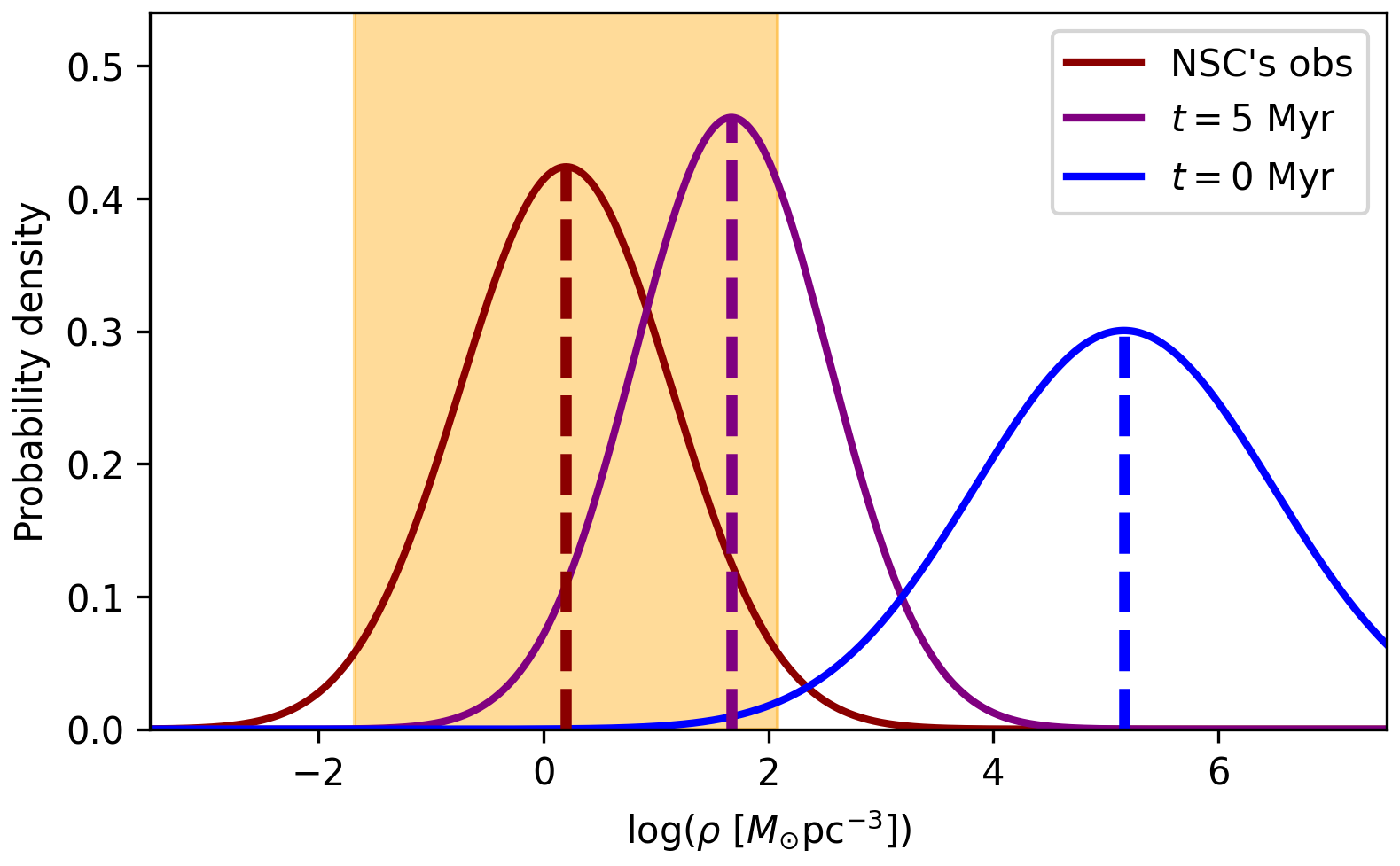}
		\caption{Distributions of $\log(\rho)$ with data from Fig.~\ref{fig:dens_mean}. The red curve is for the observational data, while blue and purple curves are for our stars-only models at $t=0~{\rm Myr}$ and $t=5~{\rm Myr}$, respectively. Vertical dashed lines indicate the respective mean values. The orange region indicates the $2\sigma$ range of the observational data.}
		\label{fig:dens_gauss}
	\end{figure}

	In Fig.~\ref{fig:dens_gauss} we show the distributions of log-densities ($\log(\rho)$). The red curve is for the observational data, while the blue and purple curves are for our models at $t=0~{\rm Myr}$ and $t=5~{\rm Myr}$, respectively. The orange region indicates the $2\sigma$ range for the observational data. At the beginning of the simulation, our models are on average $\sim 5$ orders of magnitude denser than the observational data. At the end of the simulation, the densities decrease, and the difference with respect to the observational data reduces to $\sim$1–2 orders of magnitude.
	
\end{appendix}
\end{document}